\documentclass[reprint,aps,pra,twocolumn,superscriptaddress,showpacs,longbibliography]{revtex4-1}
\usepackage{CJK}
\usepackage{bm,amsmath,amssymb,amsfonts}
\usepackage{color}
\usepackage{dcolumn}
\usepackage{graphicx}
\usepackage{float}
\usepackage{hyperref}
\hypersetup{colorlinks=true,citecolor=blue,urlcolor=blue,linkcolor=blue}
\usepackage{ulem}
\usepackage{times}
\usepackage{subfigure}

\newcommand{\finished}{{\color{red}\checkmark}}

\begin{document}
\title{Degenerate Rabi spectroscopy of the Floquet engineered optical lattice clock}
\author{Wei-Xin Liu}
\affiliation{Institute of Theoretical Physics and Department of Physics, State Key Laboratory of Quantum Optics and Quantum Optics Devices, Collaborative Innovation Center of Extreme Optics, Shanxi University, Taiyuan 030006, China}
\author{Xiao-Tong Lu}
\affiliation{Key Laboratory of Time and Frequency Primary Standards, National Time Service Center, Chinese Academy of Sciences, Xi'an 710600, China}
\affiliation{School of Astronomy and Space Science, University of Chinese Academy of Sciences, Beijing 100049, China}
\author{Ting Li}
\affiliation{Key Laboratory of Time and Frequency Primary Standards, National Time Service Center, Chinese Academy of Sciences, Xi'an 710600, China}
\affiliation{School of Astronomy and Space Science, University of Chinese Academy of Sciences, Beijing 100049, China}
\author{Xue-Feng Zhang}
\affiliation{Department of Physics, and Center of Quantum Materials and Devices, Chongqing University, Chongqing, 401331, China}
\affiliation{Chongqing Key Laboratory for Strongly Coupled Physics, Chongqing University, Chongqing, 401331, China}
\author{Hong Chang}
\affiliation{Key Laboratory of Time and Frequency Primary Standards, National Time Service Center, Chinese Academy of Sciences, Xi'an 710600, China}
\affiliation{School of Astronomy and Space Science, University of Chinese Academy of Sciences, Beijing 100049, China}
\author{Tao Wang}
\thanks{corresponding author: tauwaang@cqu.edu.cn}
\affiliation{Department of Physics, and Center of Quantum Materials and Devices, Chongqing University, Chongqing, 401331, China}
\affiliation{Chongqing Key Laboratory for Strongly Coupled Physics, Chongqing University, Chongqing, 401331, China}
\author{Wei-Dong Li}
\affiliation{Shenzhen Key Laboratory of Ultraintense Laser and Advanced Material Technology, Center for Advanced Material Diagnostic Technology, and College of Engineering Physics, Shenzhen Technology University, Shenzhen, 518118, China}
	
\begin{abstract}
Simulating physics with large SU($N$) symmetry is one of the unique advantages of Alkaline-earth atoms. Introducing periodical driving modes to the system may provide more rich SU($N$) physics that static one could not reach. However, whether the driving modes will break the SU($N$) symmetry is still lack of discussions. Here we experimentally study a Floquet engineered degenerate $^{87}$Sr optical lattice clock (OLC) by periodically shaking the lattice. With the help of Rabi spectroscopy, we find that the atoms at different Zeeman sublevels are tuned by the same driven function. Meanwhile, our experimental results suggest that uniform distribution among the sublevels will not change despite the driving. Our experimental demonstrations may pave the way to implementation of FE on tailoring the SU($N$) physics in OLC system.
\end{abstract}
\maketitle

\section{INTRODUCTION}
Alkaline-earth atoms (AEAs) (as well as the Alkaline-earth-like atom Yb) with unique atomic structures become a hot frontier in the ultra-cold atoms physics. The ultra-narrow doubly forbidden transition between ground state $^1$S$_0$ and excited state $^3$P$_0$ makes the AEAs ideal for realization of ultra-precise atomic clock. State-of-the-art optical lattice clock (OLC) using ultracold AEAs has surpassed the best $^{133}$Cs primary standards \cite{LudlowRMP2015,BloomNATURE2014,NicholsonNC2015,BothwellMetrologia2019,MGEPRL2018}. Meanwhile, owing to the strong decoupling between the nuclear spin $I$ and electronic angular momentum $J=0$ of the two lowest electronic states $^1$S$_0$ and $^3$P$_0$ (clock states), AEAs exhibit that the nuclear spin is independent of both interatomic collision and trapping potential. Hence, it directly leads to the SU($N \leq 2I+1$) symmetry emerging in the AEAs \cite{WCJMPLb2006,GorshkovNP2010,PaganoNP2014,ScazzaNP2014,MiguelRPP2014,ZhangSCIENCE2014,PaganoNP2014,BeverlandPRA2016,CGPRA2016,BanerjeePRL2013,PerlinNJP2019,HeJPB2019,ChoudhuryPRA2019,WCJPRL2003,WuNP2012,WuPhysics2010,CGandWCJ2021}. AEAs possessing high-dimensional symmetries with large $N$ (e.g. 10 for $^{87}$Sr) are predicted to simulate the high-energy lattice gauge theories \cite{BDPRL2013}, but also a unique platform for investigating a variety of many-body phases \cite{HCPRL2004,HMPRL2009,HMPRB2011,CMANJP2009}. Recently, more and more experiments support the existence of the SU($N$) symmetry \cite{StellmerPRA2011,TSNP2012,ZhangSCIENCE2014,PaganoNP2014,ScazzaNP2014,CGPRL2014}.	

On the other hand, increasing efforts have been devoted to manipulating ultracold atoms using time-periodic modulations. Focusing on the optical lattice, the modulation can  provide an extremely clean system with highly controllability in time-dependent fashion. This coherent manipulation of quantum system is known as the Floquet engineering (FE) \cite{EckardtRMP2017}. FE of the ultracold atoms in optical lattice shows a high potential for simulating and studying a wide variety of condensed-matter systems, and even some models in high-energy physics. It has achieved many successes such as the dynamic control of insulator-superfluid quantum phase transition \cite{AZPRL2009}, the realization of topological band structure \cite{CooperPMP2019,JotzuNATURE2014,AidelsburgerNP2015}, and the creation of artificial gauge field \cite{AidelsburgerPRL2013,StruckPRL2012,StruckNATURE2013,MiyakePRL2013}. Thus, introducing FE into SU($N$) physics of the AEAs Fermi gases becomes more attractive, such as renormalizing the tunneling of SU($N$) Hubbard model or generating exotic SU($N$) phase \cite{ZhangNRP2020}. However, the compatibility of FE with the SU($N$) symmetry needs to be addressed firstly before pursuing these amazing prospects. Specifically, will atoms at different sublevels be tuned by the same driving function, and will uniform distribution among sublevels be changed by periodic driving?

\begin{figure}[t]
	\centering
	\includegraphics[width=1\linewidth]{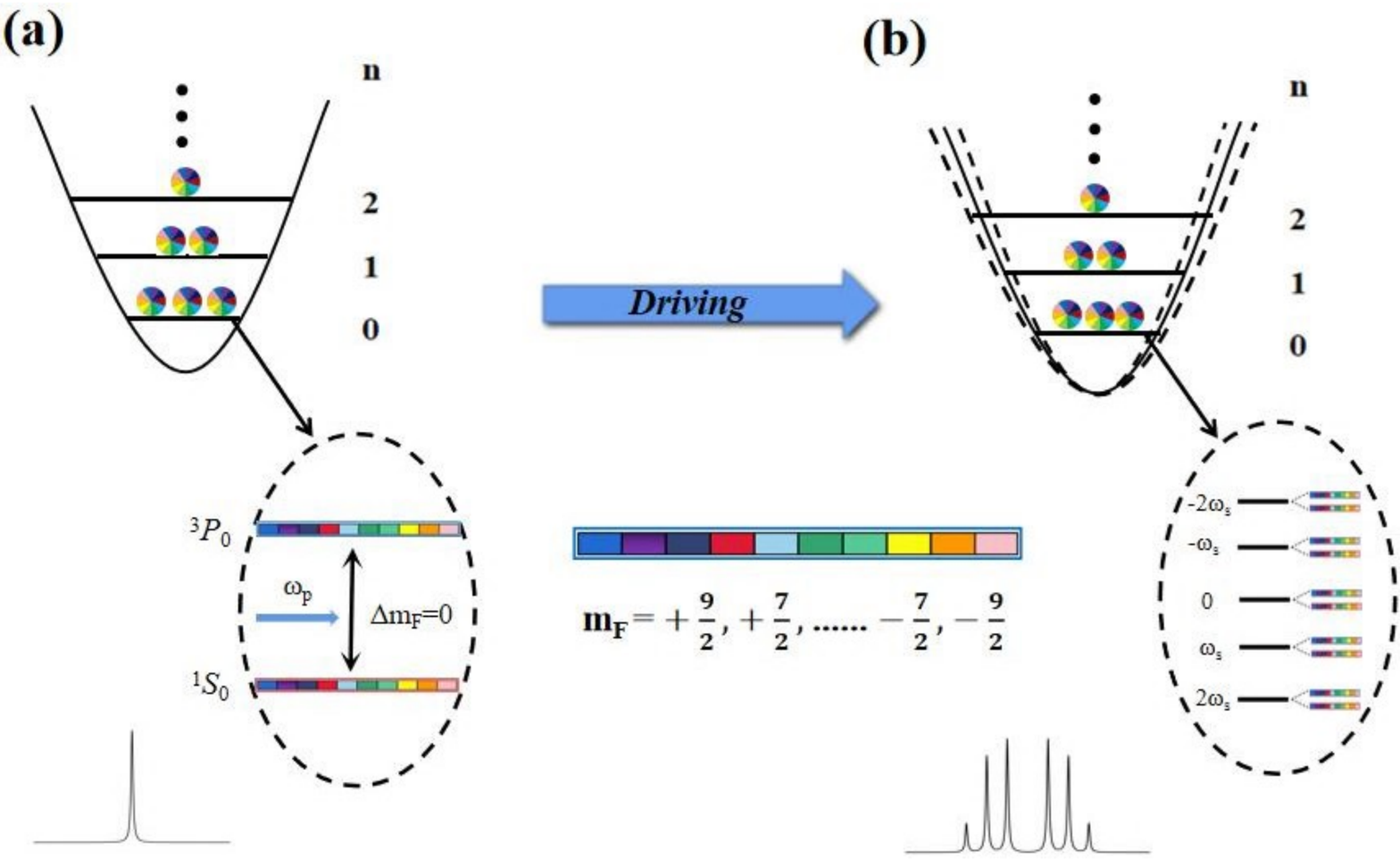}
	\caption{\finished Floquet engineered degenerate optical lattice clock. In the absence of magnetic field, both clock states have ten degenerate sublevels (denoted by different colors \cite{ZhangSCIENCE2014}). (a) Without driven, the atoms follow the Boltzmann distribution in the external states. After the $\pi$ clock transitions, the atoms can hop between two clock states without changing their Zeeman sublevels $\Delta m_F=0$. (b) Under the lattice driving pattern, the carrier peak in Rabi spectroscopy can split to several Floquet sidebands. Each Floquet sideband corresponds to the $\pi$ clock transitions in each Floquet energy level which can still keep the SU($N$) symmetry.}
	\label{Fig0}
\end{figure}

In this manuscript, we experimentally realized a Floquet engineered degenerate clock transition in one-dimensional (1D) $^{87}$Sr OLC, and also demonstrate that FE not apparently change the distribution of atoms among degenerate energy levels with the help of Rabi spectroscopy. As shown in Fig. \ref{Fig0}, the fermionic $^{87}$Sr has a nuclear spin of $I=9/2$, therefore, both the two clock states have tenfold degeneracy ($2F+1=10$, where $F=I+J$ is the total atomic angular momentum) in the absence of magnetic field, corresponding to the ten Zeeman sublevels from $m_F=-9/2$ to $m_F=9/2$ ($m_F$ is the magnetic quantum number of total angular momentum). Then, all the ten degenerate sublevels can be Floquet engineered by modulating the lattice laser around the ``magic" wavelength in a nearly zero magnetic field. Under this driving pattern, the internal dynamics of $\pi$ transition in each $m_F$-sublevel is governed by the time-dependent Landau-Zener-St\"{u}ckelberg-Majorana (LZSM) Hamiltonian due to the Doppler effect \cite{YinCPL2021,Wangprl2021,LZPRA2021}. The total clock transition probability can be theoretically calculated, basing on the resolved Floquet sideband approximation (RFSA). After extracting The experimental parameters in the nondriven case, especially the bare Rabi frequencies of all the sublevels, we estimate the residual stray magnetic field. At last, we measure the Floquet degenerate Rabi spectrum and the Rabi oscillation, and find the periodic modulation will not break the SU($N$) symmetry.

The manuscript is organized as follows. In Sec. II, the experimental setup is introduced. In Sec. III, the model of degenerate driven system are calculated. In Sec. IV, we discuss how to determine the experimental parameters. In Sec. V, we give an analysis of the SU($N$) symmetry and show the degenerate Floquet Rabi spectroscopy. Sec. VI includes the conclusions and outlook.

\section{EXPERIMENTAL SETUP}
Approximately $10^4$ $^{87}$Sr atoms are cooled to about $3$ $\mu$K by standard laser cooling techniques and trapped in a 1D optical lattice in the Lamb-Dicke region, where the motion and the photon recoil momentum of the atom will not broaden the clock transition spectra \cite{DickePR1953,MukaiyamaPRL2003,TakamotoPRL2003}. The 1D optical lattice consists of two counter-propagating laser beams at the ``magic'' wavelength $\lambda_L =813.43$ nm \cite{TakamotoPRL2003}, where the AC Stark frequency shifts of the two clock states $(5s^2)^1$S$_0$ ($|g\rangle$) and $(5s5p)^3$P$_0$ ($|e\rangle$) are equal. One incident lattice laser beam, with a linear polarization along the direction of gravity, is focused onto the center of magneto-optical trap (CMOT). After a high-reflection mirror, its retro-reflected laser beam is also focused onto the CMOT and forms a standing wave with incident laser. The lattice laser (TOPTICA photonics AG, Munich, Germany) has a power of 300 mW and beam waist $w_0 \simeq 50$ $\mu$m around the CMOT. Because of the large trapping depth, the tunneling between lattice sites can be ignored so that the system could be taken as a series of independent harmonic traps. Then, the eigenstates of each harmonic trap are labeled as the external states $|\vec{n}\rangle$ or $|n_r,n_z\rangle$ which $n_{r(z)}$ corresponding to transverse (longitudinal) direction of the optical lattice potential \cite{YinCPL2021,ShallowYin2021}.

The clock transition $^1$S$_0 \leftrightarrow ^3$P$_0$ is interrogated by the clock laser (DL pro, TOPTICA Photonics AG, Munich, German) with a wavelength of $698$ nm propagating collinear with the lattice laser. The clock laser is divided into two parts by a beam splitter (BS). Ninety percent of the clock laser directly passes the BS, is collimated by lens assembly and then completely overlaps with the lattice laser. Ten percent is reflected by BS and then entered into a 10-cm-long ultralow-expansion ultra-stable cavity with a finesses of 400000. The clock laser beam has the same polarization direction as the lattice laser beam, and its beam waist is $2$ mm around the CMOT. The full width at half maximum of clock laser is narrowed to $1$ Hz after Pound-Drever-Hall locking, and the short term stability of the clock laser is $1\times10^{-15}$ at 1 s \cite{WYBCPB2018}. The natural lifetime of $^3$P$_0$ is about $120$ s and the duration of the measurement processes does not exceed 1s, so we can ignore the spontaneous emission during the clock transition detection which is conducted by the method of ``electronic shelved" \cite{NWPRL1986,TMPRL2003}.

By changing the currents of three-dimensional compensation coils (TDCCs), we fine-tune the magnetic field around the atoms approaching to zero as possible as we can, in order to avoid breaking the SU($N$) symmetry while Floquet engineering the degenerate system. Then we apply a periodic sinusoidal modulation to the piezoelectric transducer (PZT) adhered to the grating to periodically change the cavity length of the lattice laser. Under this modulation, the lattice frequency can be expressed as $\omega_L(t)=\bar{\omega}_L + \omega_a\sin(\omega_st)$, where $\bar{\omega}_L = 2\pi c/\lambda_L$ is the central lattice frequency at ``magic" wavelength $\lambda_L$, $\omega_{a(s)}$ is the driving amplitude (frequency) which is typically several hundreds of MHz (Hz) in our experiment. The degenerate driving spectra can be obtained by scanning the frequency of the clock laser with help of an acousto-optic modulator in each clock detection cycle. The power of clock laser is set to be 220 nW, so that the effect of saturation-broadening can be neglected \cite{TMPRL2003}.

\section{THE MODEL}
Under the lattice laser frequency modulation, the atoms at different sublevels may pick up a sublevel-dependent velocity in the co-moving frame of lattice \cite{YinCPL2021}. 
Due to the Doppler effect, the dynamics of the atoms at certain $m_F$ in external energy level $|\vec{n}\rangle$ is governed by the time-dependent LZSM Hamiltonian in the lattice co-moving frame \cite{YinCPL2021,Wangprl2021,LZPRA2021}
\begin{equation}
\hat{H}_{\vec{n}}^{m_F}(t) = \frac{\hbar}{2}[\delta_{m_F}+A_{m_F}\omega_s\cos(\omega_st)]\hat{\sigma}_z + \frac{\hbar}{2}g_{\vec{n}}^{m_F}\hat{\sigma}_x.    \label{0}%
\end{equation}
Here,  $\hbar$ is the reduced Planck's constant, and the detuning $\delta_{m_F} = \delta+\Delta\omega_0^{m_F}$ includes the bare clock laser detuning $\delta = \omega_p-\omega_0$ but also the $\pi$ transition frequency shift for $m_F$-sublevel $\Delta\omega_0^{m_F}$, because there is still an unavoidable small residual magnetic field in the experiment. $\omega_0$ and $\omega_p$ are the bare clock transition frequency and clock laser frequency, respectively. $A_{m_F}$ is the sublevel-dependent renormalized driving amplitude, and $g_{\vec{n}}^{m_F}$ is the coupling strength of $m_F$-sublevel in the external state $|\vec{n}\rangle$ \cite{YinCPL2021,BlattPRA2009}.

The $\pi$ transition frequency shift is derived from the AC stark shift caused by the lattice laser acting on the Zeeman sublevels and the possible residual stray magnetic field. In the present of lattice potential with trapping depth $U_0$ at a weak magnetic field $B$, the $\pi$ transition frequency shift between $m_F$-sublevel can be expressed as \cite{BoydPRA2007,ShiPRA2015,WestergaardPRL2011}
\begin{align}
\Delta\omega_0^{m_F} & = (\Delta \kappa^S + \Delta \kappa^V m_F \xi \vec{e}_k \cdot \vec{e}_B + \Delta \kappa^T\beta) U_0 \nonumber\\
                     & + \delta g m_F \mu_0 B ,                                                                                            \label{1}
\end{align}
where $\Delta\kappa^S$, $\Delta\kappa^V$, and $\Delta\kappa^T$ are the coefficients due to differential scalar, vector and tensor polarizabilities between $^1$S$_0$ and $^3$P$_0$, respectively. $\xi$ is the degree of ellipticity of the lattice light field, where $\xi=0$ ($\pm1$) represents perfect linear (circular) polarization. $\vec{e}_B$ and $\vec{e}_k$ are unitary vectors along the quantization axis and the lattice wave vector, respectively. The coefficient $\beta$ is equal to $(3\cos^2\phi-1)[3m_F^2-F(F+1)]$ with $\phi$ the angle between the linear polarization direction of the lattice laser and the quantization axis. The last term in Eq. (\ref{1}) is the first order Zeeman shift in which $\delta g$ is the differential Land\'{e} factor between the two clock states due to the hyperfine interaction on the excited state $^3$P$_0$ \cite{BoydPRA2007,LurioPR1962}. $\mu_0$ is equal to $\mu_B/h$ with $\mu_B$ the Bohr magneton and $h$ the Planck's constant. The hyperpolarizability effect ($\propto U_0^2$) and the second order Zeeman shift have been ignored in Eq. (\ref{1}) as they are negligible at the level of 1 mHz \cite{BruschPRL2006,WestergaardPRL2011,BaillardOL2007,UshijimaPRL2018}.

First, the effect of scalar shift can be omitted, because it is independent of $m_F$ while the lattice laser frequency varying around the ``magic" wavelength. Second, the effect of periodic driving on the external potential could also be neglected due to the small driving amplitude \cite{YinCPL2021}. Third, considering the lattice field is linear polarization along the quantization axis, we can directly obtain $\xi \simeq 0$, $\vec{e}_k \cdot \vec{e}_B \simeq 0$, and $\phi \simeq 0$, so that the vector shift term is also omitted. Last, according to the Ref. \cite{WestergaardPRL2011}, the tensor shift coefficient is about $-0.06$ mHz/$E_R$. Because $U_0\simeq94 E_R$ in our experiment (see Appendix), the tensor shift is less than $1$ Hz and can also be neglected. Thus, the main frequency shift is only the last term which depends on $m_F$ and the magnetic field B with $\delta g\mu_0 = -108.4(4)$ Hz/G \cite{BoydPRA2007}. Then, the $\pi$ transition frequency shift can be simplified as $\Delta\omega_0^{m_F}  = - 108.4 m_F B$.

The population of atoms is evenly distributed among ten Zeeman sublevels after the second stage of cooling \cite{BoydPRL2007}. In the absence of driving, SU($N=10$) symmetry implies the number of atoms in each of ten sublevels is conserved \cite{ZhangSCIENCE2014,StellmerPRA2011}.
The dynamics of atom distribution due to the FE is still an open question, so here we can introduce a sublevel-dependent distribution $N_{m_F}/N_0$ during the FE, where $N_0$ is the total atom number, and it will be determined from the experiments later.
The Floquet spectrum can be treated as the summation of contributions from all the independent degenerated sublevels, so the total excitation probability in terms of the Zeeman sublevels can be obtained by solving the Hamiltonian Eq. (\ref{0}) with the RFSA \cite{YinCPL2021}
\begin{equation}
P_e^{\rm Tol}=\sum_{m_F}P_e^{m_F} ,                 \label{2}
\end{equation}
where
\begin{equation}
P_e^{m_F}\!=\!\sum_{\vec{n},k}q(\vec{n})\!\left[\frac{N_{m_F}g_{\vec{n}}^{m_F}J_k(A_{m_F})}{N_0 g_{\rm eff}^{k,m_F}}\right]^{2}\!\sin^{2}\left(\frac{g_{\rm eff}^{k,m_F}}{2}t\right) \label{22}
\end{equation}
is the excitation probability of $m_F$-sublevel, and
\begin{equation}
g_{\rm eff}^{k,m_F} = \sqrt{(\delta_{m_F}-k\omega_s)^{2} + [g_{\vec{n}}^{m_F}J_k(A_{m_F})]^2} \label{3}
\end{equation}
is the effective Rabi frequency for $k$th-order sideband of the $m_F$-sublevel, $J_k(.)$ is the $k$th-order Bessel function of the first kind, $q(\vec{n})$ is the Boltzmann factor. The coupling strength can be explicitly written as $g_{\vec{n}}^{m_F}=g_{m_F}e^{-\eta_z^2/2}e^{-\eta_r^2/2}L_{n_z}(\eta_z^2)L_{n_r}(\eta_r^2)$, where $\eta_z=\sqrt{h/(2m\nu_z)}/\lambda_p$, $\eta_r=\sqrt{h/(2m\nu_r)}\Delta\theta/\lambda_p$ are the Lamb-Dicke parameters, $\nu_{z(r)}$ is the longitudinal (transverse) trap frequency, $\lambda_p$ is the clock laser wavelength, $\Delta\theta$ is the residual misalignment angle between lattice and probe axis, and $m$ is the mass of the atom \cite{YinCPL2021,BlattPRA2009}. In order to theoretically obtain the excitation probability \cite{YinCPL2021,Wangprl2021,ShallowYin2021}, the experimental parameters should be determined firstly.

\section{Determination of the experimental parameters}
In the nondriven case, the SU($N$) symmetry is not broken, so $N_{m_F}/N_0=1/N$. Then, we can determine the experimental parameters from the degenerate Rabi spectrum in the nondriven system. In Table \ref{Table1}, we list some determined experimental parameters which require the same methods as the polarized case (see Ref. \cite{YinCPL2021} and Appendix). Beside that, we needs to determine the bare Rabi frequencies $g_{m_F}$ of all the sublevels as well as the small residual magnetic field $B$ without driving.
\begin{table}[h!]
	\begin{center}
		\begin{tabular}{|c|c|c|c|c|c|}
			\hline
			parameter & value & parameter & value & parameter & value\\
			\hline
			$T_z$   &  $3.5 \mu$K & $\nu_z$ &  $66.8$ kHz & $N_z$  &  $5$ \\
			\hline
			$T_r$   &  $4.0 \mu$K & $\nu_r$ &  $250$ Hz   & $N_r$  & $1336$ \\
			\hline
		\end{tabular}
		\caption{The determinated experimental parameters.}\label{Table1}
	\end{center}
\end{table}
\begin{figure}[t]
	\centering
	\includegraphics[width=1\linewidth]{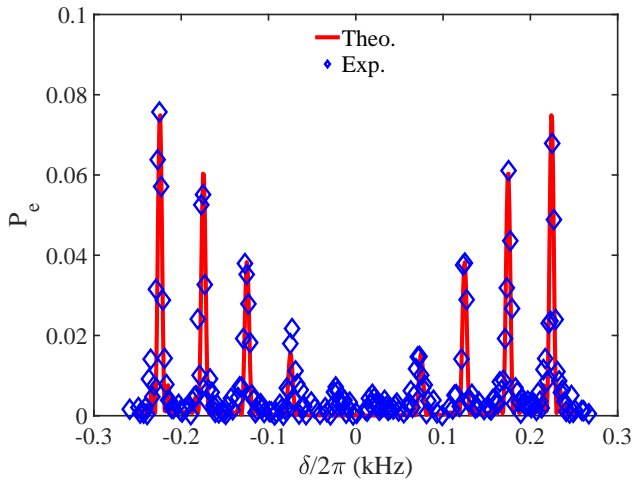}
	\caption{The Zeeman spectrum in the presence of a $460$ mG magnetic field at $150$ ms clock laser interrogation time. These peaks from left to right correspond to the $\pi$ transition for $m_F=-9/2$ to $m_F=+9/2$ sublevel, respectively. }%
	\label{Fig1}%
\end{figure}

\begin{figure}[t]
	\centering
	\includegraphics[width=0.5\linewidth]{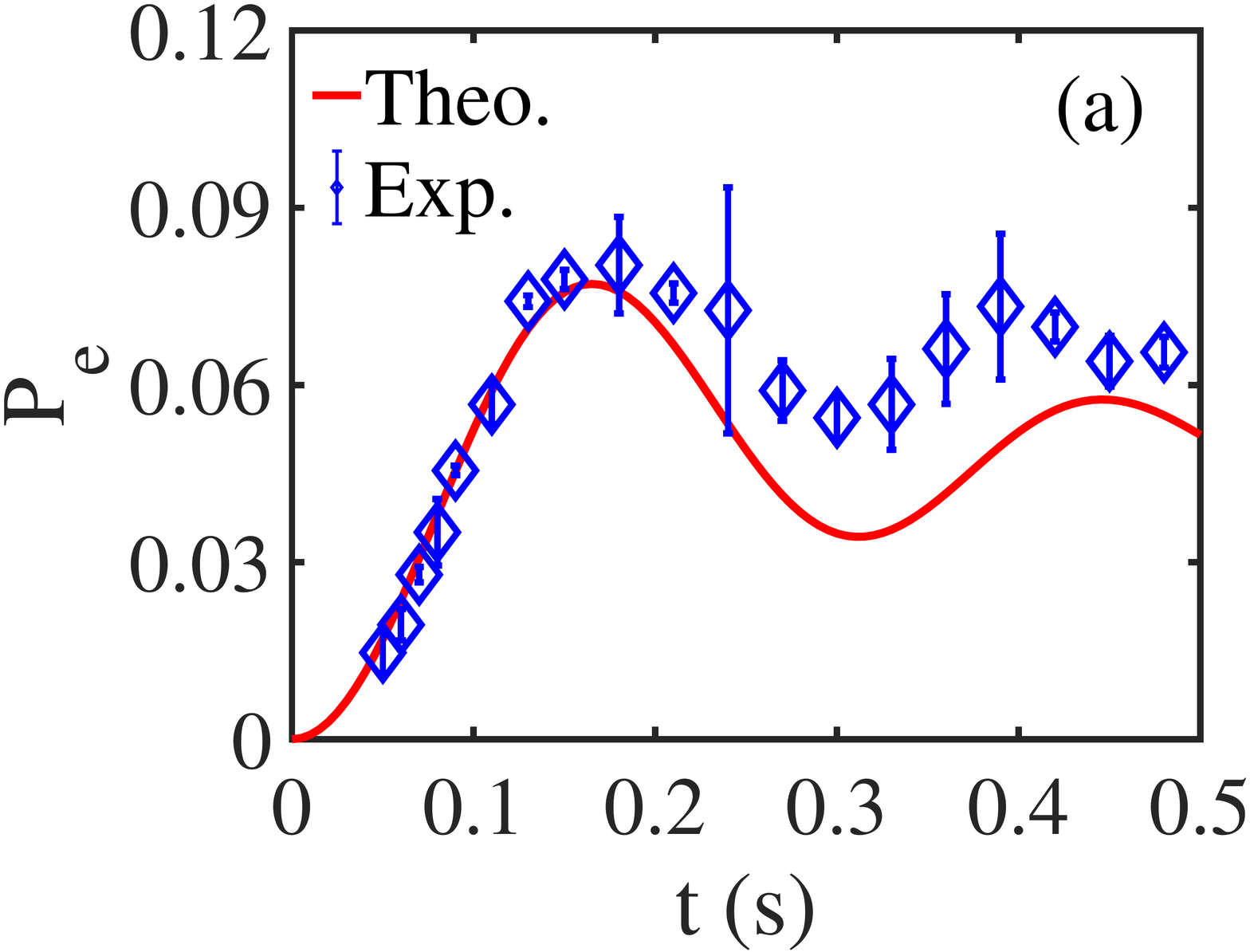}\hfill
	\includegraphics[width=0.5\linewidth]{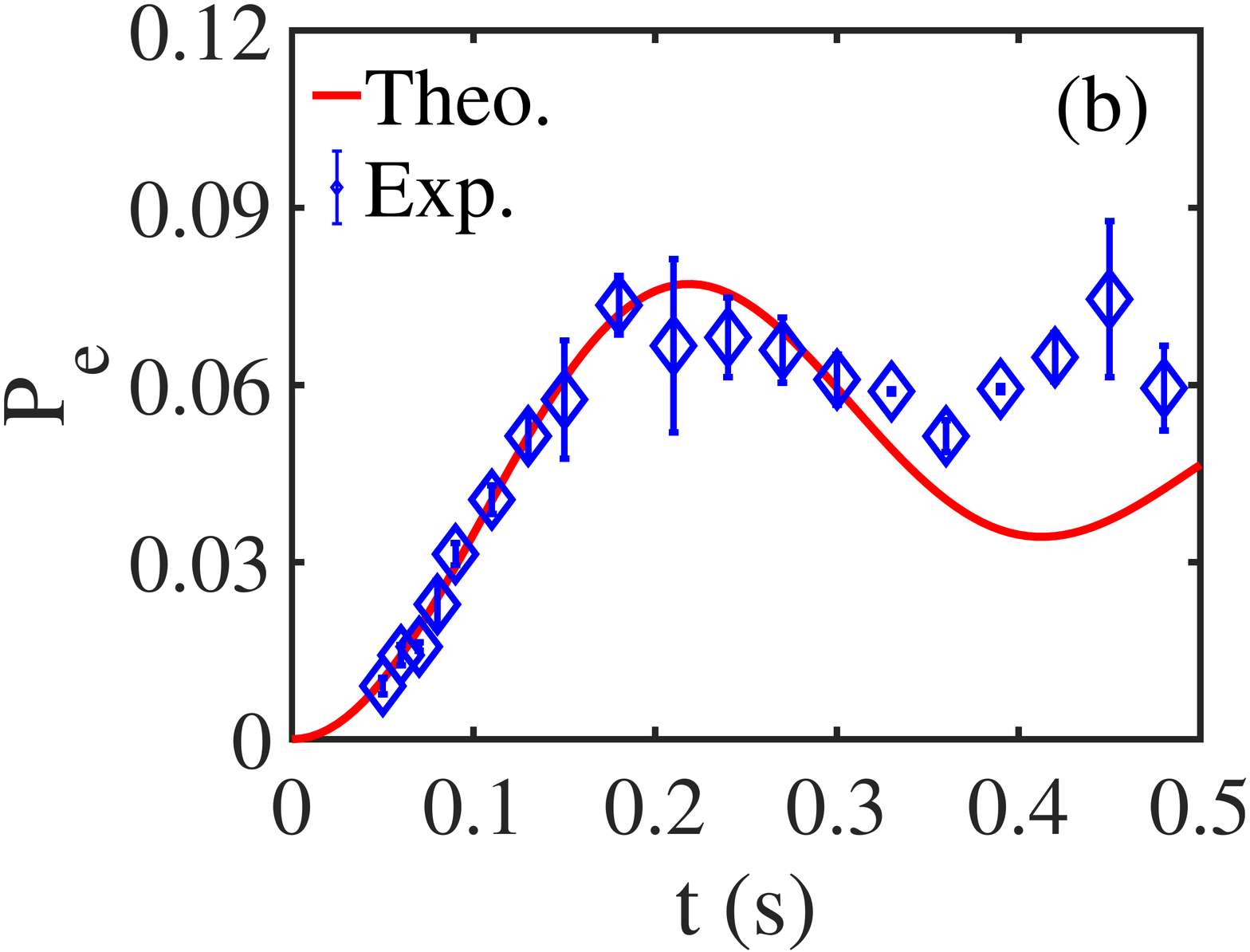}\\
	\includegraphics[width=0.5\linewidth]{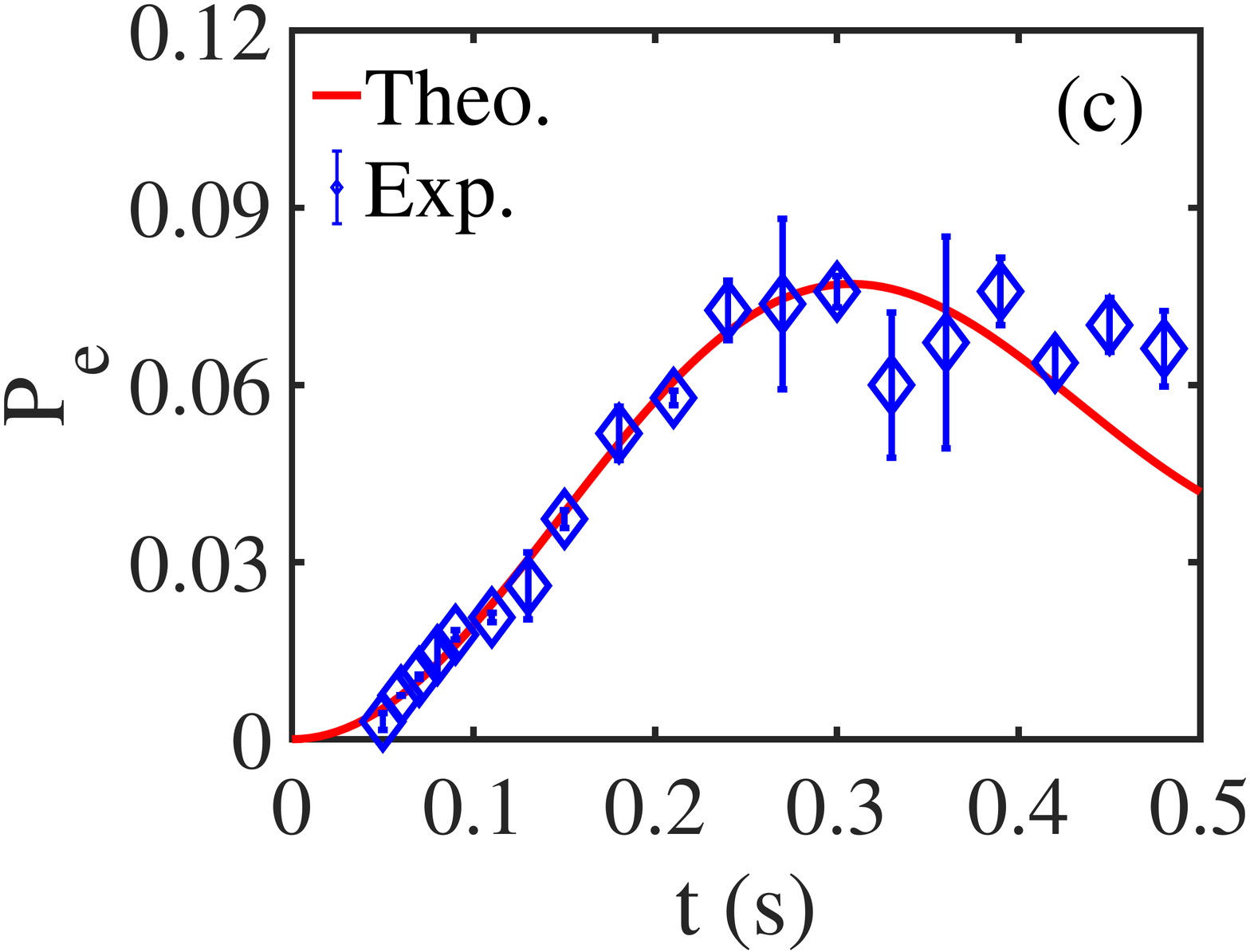}\hfill
	\includegraphics[width=0.5\linewidth]{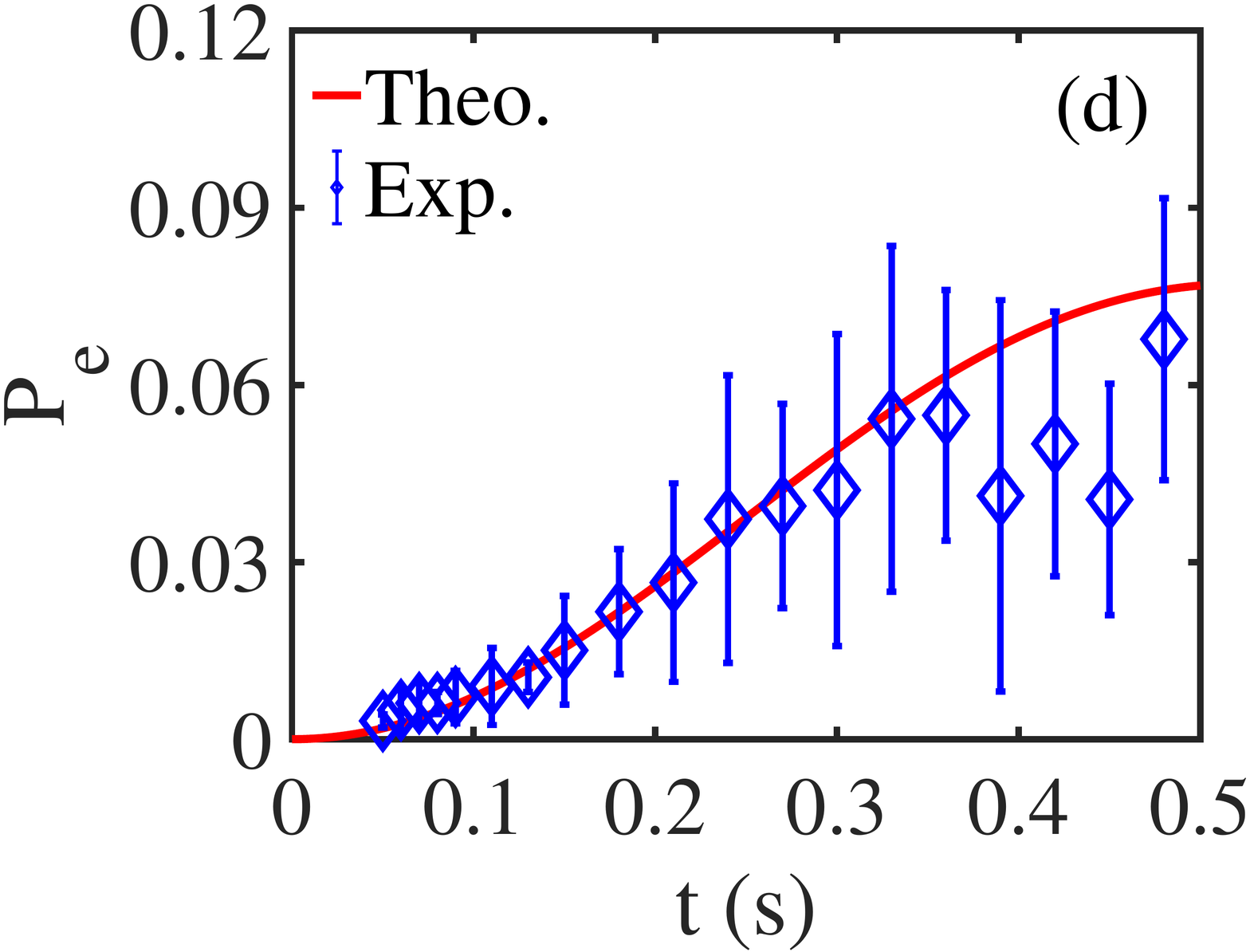}
	\caption{Rabi oscillations of different Zeeman sublevels at (a) $|m_F|=9/2$, (b) $|m_F|=7/2$, (c) $|m_F|=5/2$ and (d) $|m_F|=3/2$. The experimental parameters are the same as Fig. \ref{Fig1}}%
	\label{Fig2}%
\end{figure}

First, we use the nondriven Zeeman spectrum to extract the Rabi frequency of each sublevel. In order to split the Zeeman sublevels, the currents of TDCCs are adjusted to as large as $460$ mG along the direction of the gravity. As shown in Fig. \ref{Fig1}, the Zeeman spectrum presents eight obvious peaks with intervals close to the theoretical prediction $\delta\Delta\omega_0^{m_F}=108.4B\approx$ 50Hz, which is much larger than linewidths of spectrum with a few Hz. Due to the first order Zeeman shift, these peaks correspond to the Zeeman sublevels except $m_F=\pm1/2$ which have quite weak excitation probability. By scanning the clock laser frequency at a fixed clock laser interrogation time and clock laser power ($220$ nW), we get a set of Zeeman spectra in Fig. \ref{Fig1}. Then we can extract excitation fraction for each sublevel from the Zeeman spectra under different interrogation times and get the Rabi oscillations of the sublevels as shown in Fig. \ref{Fig2}(a)-(d) corresponding $|m_F|=9/2,7/2,5/2,3/2$, respectively. In order to improve the experimental data, here we take average of sublevels with same $|m_F|$ due to positive-negative or $\mathbb{Z}_2$ symmetry. In addition, the Rabi oscillation of sublevel $|m_F| = 1/2$ is not shown here because it is too small, and we also ignore its effect on the Floquet spectra in the following. Then by fitting the experimental data (before $200$ ms interrogated time) for each Zeeman sublevel with Eq. (\ref{22}) taking $A_{m_F}=0$, $\omega_s=0$, and $N_{m_F}/N_0=1/N$, we can get the Rabi frequency $g_{|m_F|}$ in each sublevel and the misalignment angle $\Delta\theta$ which should be same for all sublevels \cite{YinCPL2021,BlattPRA2009}. Then, the fitting results are $g_{\pm9/2}/2\pi = 4.1$ Hz, $g_{\pm7/2}/2\pi = 3.1$ Hz, $g_{\pm5/2}/2\pi = 2.2$ Hz, $g_{\pm3/2}/2\pi = 1.3$ Hz at a fixed average misalignment angle $\overline{\Delta \theta} = 8$ mrad.
With the help of these Rabi frequencies, we show the theoretical Zeeman spectrum with Eq. (\ref{2}) by setting $A_{m_F}=0$, $\omega_s=0$, $B=460$ mG and $t=150$ ms in Fig. \ref{Fig1}, and it well agrees with the experimental data.

\begin{figure}[b]
\centering
\includegraphics[width=1\linewidth]{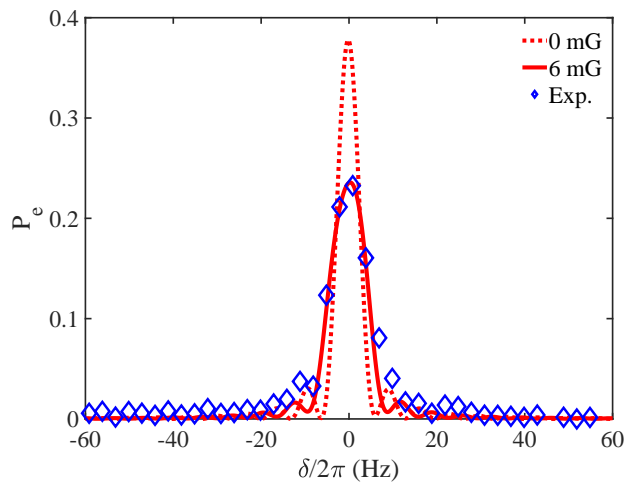}\\
\caption{The degenerate narrow spectrum at $150$ ms clock laser interrogation time. The blue diamonds are the experimental data. The solid and dashed lines are the theoretical results with magnetic field $B=6$ mG and zero, respectively.}%
\label{Fig3}%
\end{figure}

When we tune the currents of TDCCs to decrease the magnetic field, the splitting will become less obvious and finally all the peaks will merge into single broad peak. Although we can narrow the linewidth by further fine-tuning the TDCCs, the exact zero magnetic field still can not be achieved. In order to further determine the strength of residual magnetic field, we keep the power of the clock laser at $220$ nW and scan the clock laser frequency under the $150$ ms interrogation time, and the Rabi spectrum without driven is shown in Fig. \ref{Fig3}. This narrow spectrum can be taken as the summation of contributions from all the sublevels with uniform atom distribution.
Thus we can determine the stray magnetic field $B$ by fitting the experimental data with Eq. (\ref{2}) taking $A_{m_F}=0$, $\omega_s=0$ and the Rabi frequencies of sublevels determined before. As shown in Fig. \ref{Fig3}, in contrast with the zero magnetic field, the best-fitting stray magnetic field $B$ is equal to 6mG.

\section{Analysis of the SU($N$) Symmetry }
The experimental parameters obtained in the nondriven case are assumed to be unchanged while the periodic modulation is switched on, except the atom distribution $N_{m_F}/N_0$ due to the possible atom transport among the sublevels. Meanwhile, the renormalized driving amplitude $A_{m_F}$ could also be different, because the sublevels may have different reaction to the periodic modulation.

\begin{figure}[t]
	\centering
	\includegraphics[width=0.5\linewidth]{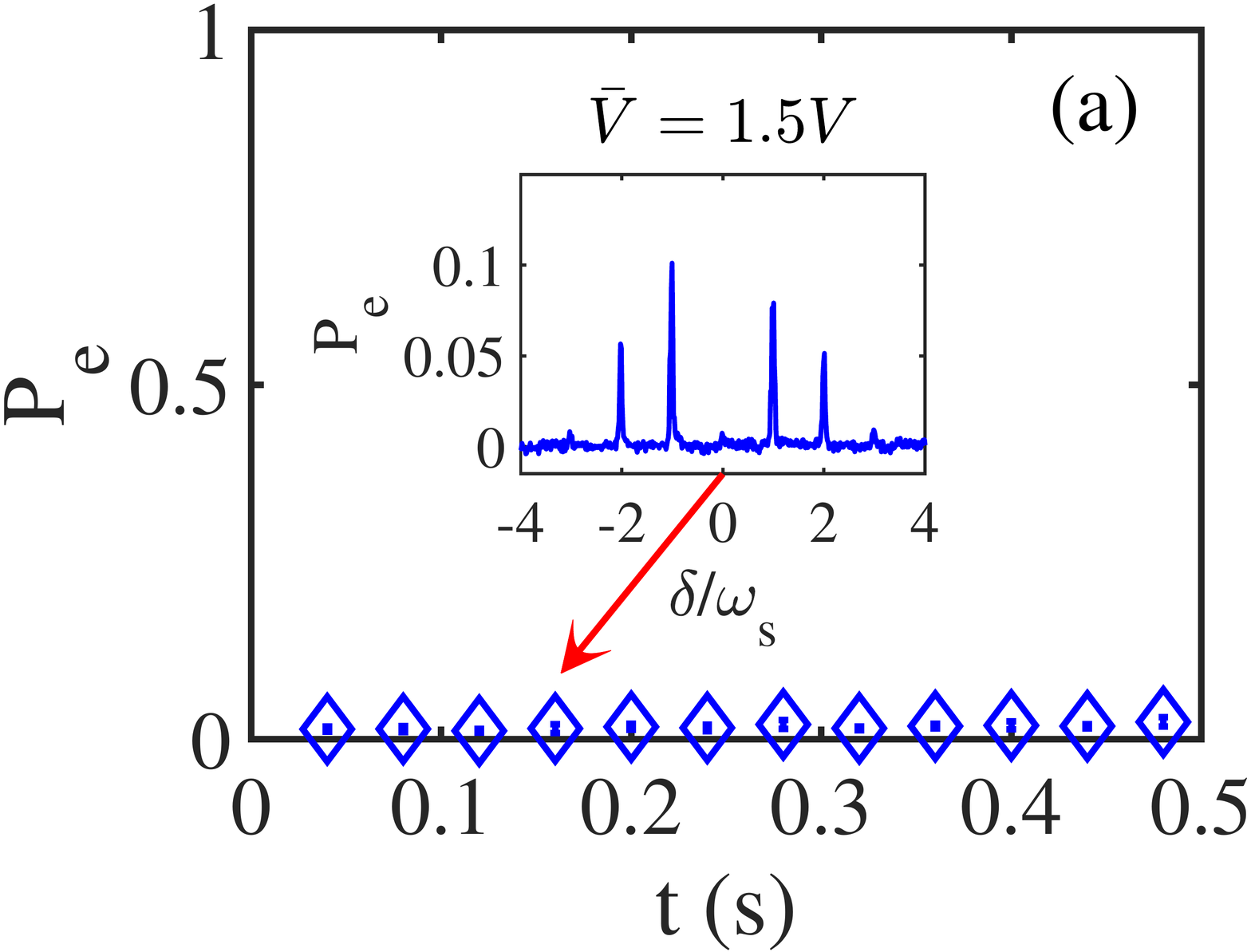}\hfill
	\includegraphics[width=0.5\linewidth]{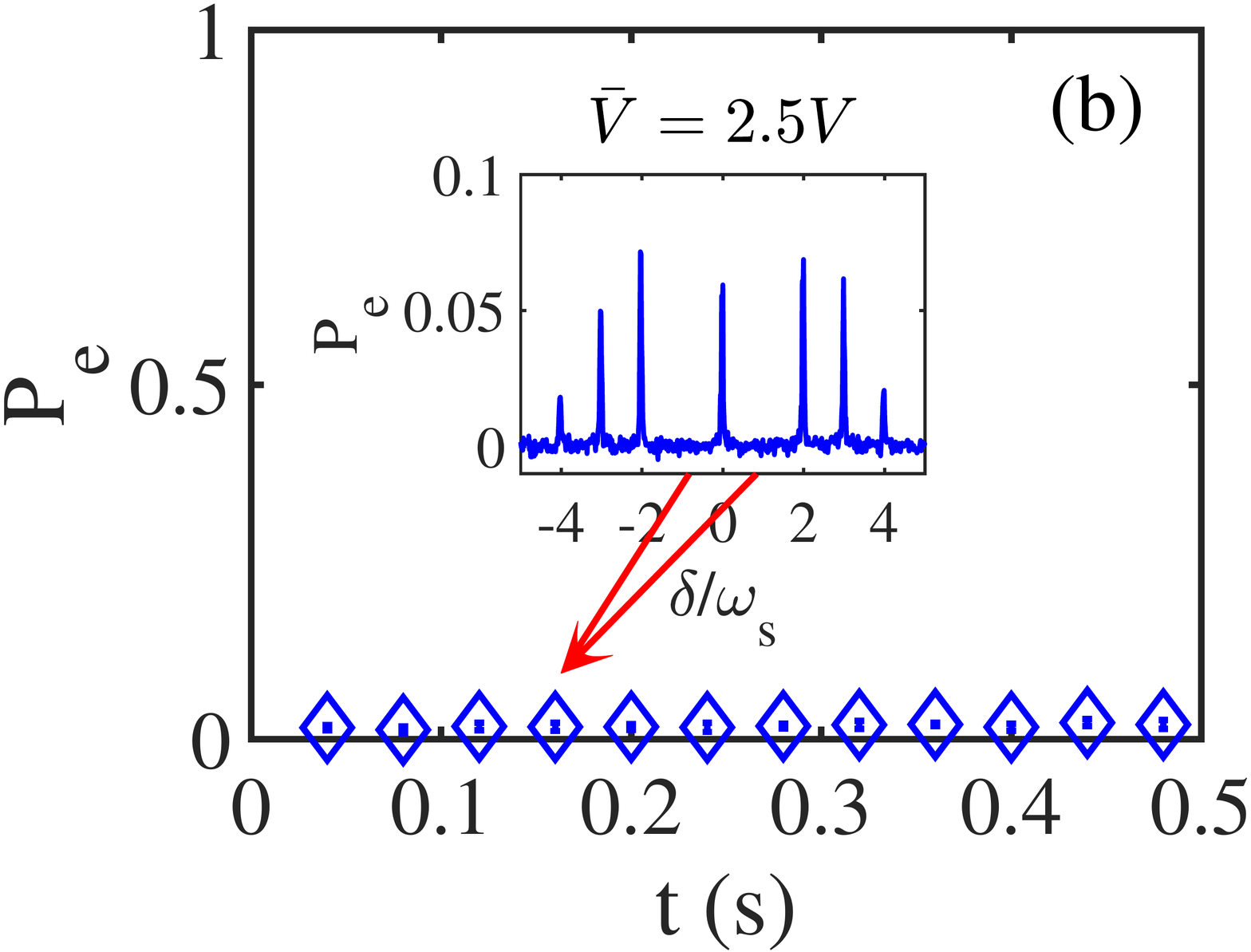}\\
	\caption{Rabi oscillations of (a) the zeroth Floquet side-band at $\bar{V}=1.5V$ and (b) the first Floquet side-band at $\bar{V}=2.5V$. The insets show the corresponding Rabi spectroscopy, respectively.}%
	\label{Fig4}%
\end{figure}

First, it is easy to verify whether the renormalized driving amplitude is sublevel dependent. If they are different, the corresponding Bessel function $J_k(A_{m_F})$ can not be fine-tuned to zero for all the sublevels, so that the excitation population of the $k$-th order Floquet sideband can not be totally suppressed to zero. In the experiment, the renormalized driving amplitude $A_{m_F}$ can be tuned by changing the voltages added to the PZT $\bar{V}$ \cite{YinCPL2021}. As shown in the insets of Fig. \ref{Fig4}, the zeroth (first) Floquet sideband of the Floquet Rabi spectroscopy is suppressed lower than the background noise at $\bar{V}=1.5V$ ($\bar{V}=2.5V$). To rule out the accidental case, we also measure the Rabi oscillation of these suppressed Floquet sidebands in Fig. \ref{Fig4}. The experimental data clearly shows that the excitation populations of both zeroth Floquet sideband at $\bar{V}=1.5V$ and first one at $\bar{V}=2.5V$ are almost zero up to $500$ms. They straightforwardly demonstrate all the renormalized driving amplitude $A_{m_F}$ are fine-tuned to the same certain suppressing value, and it will be strange that $A_{m_F}$ are sublevel dependent at other voltage values. Thus, we set the renormalized driving amplitude of all sublevels to be the same value $A_{m_F}=A(\bar{V})$ at voltage $\bar{V}$.

\begin{figure}[t]
	\centering
	\includegraphics[width=0.5\linewidth]{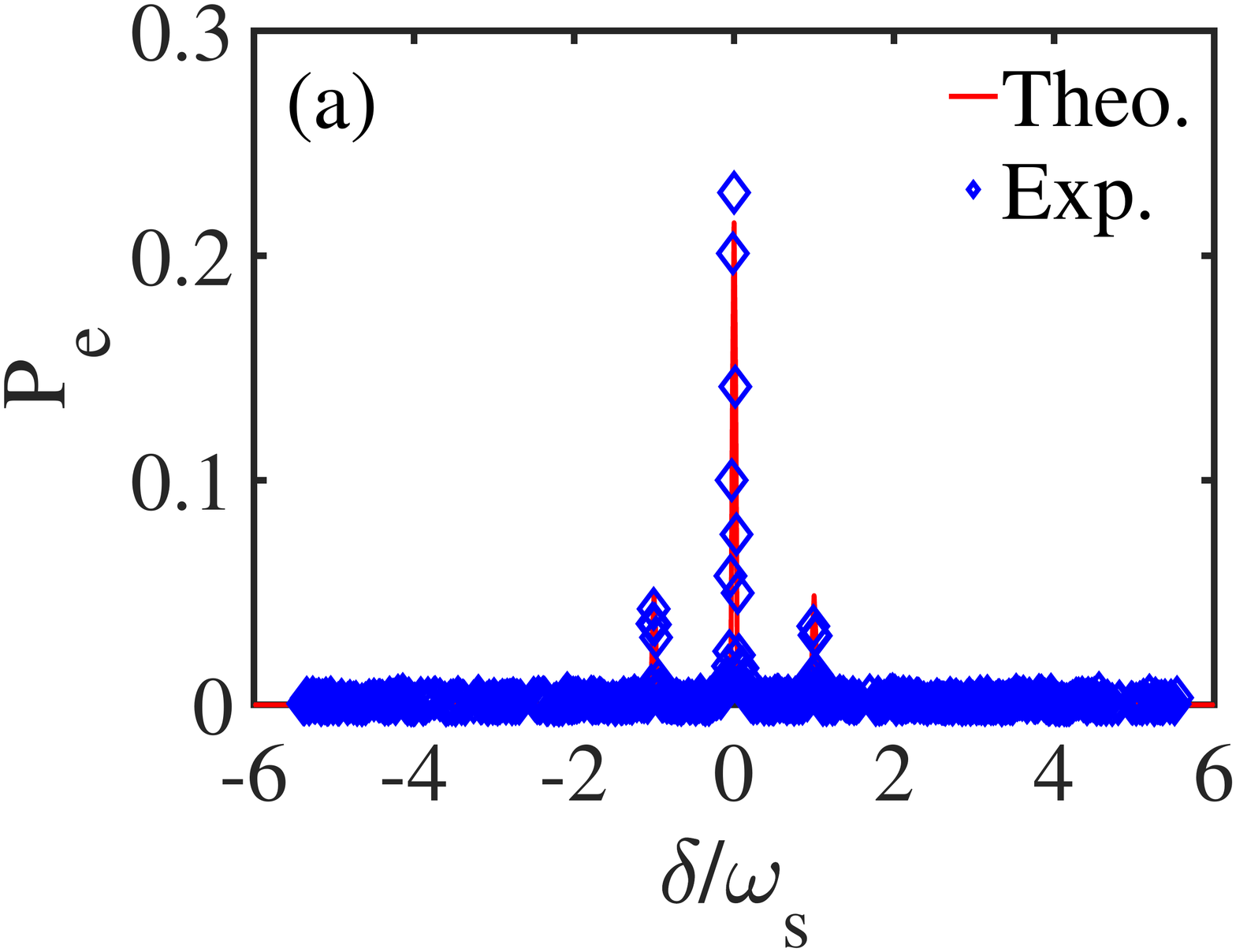}\hfill
	\includegraphics[width=0.5\linewidth]{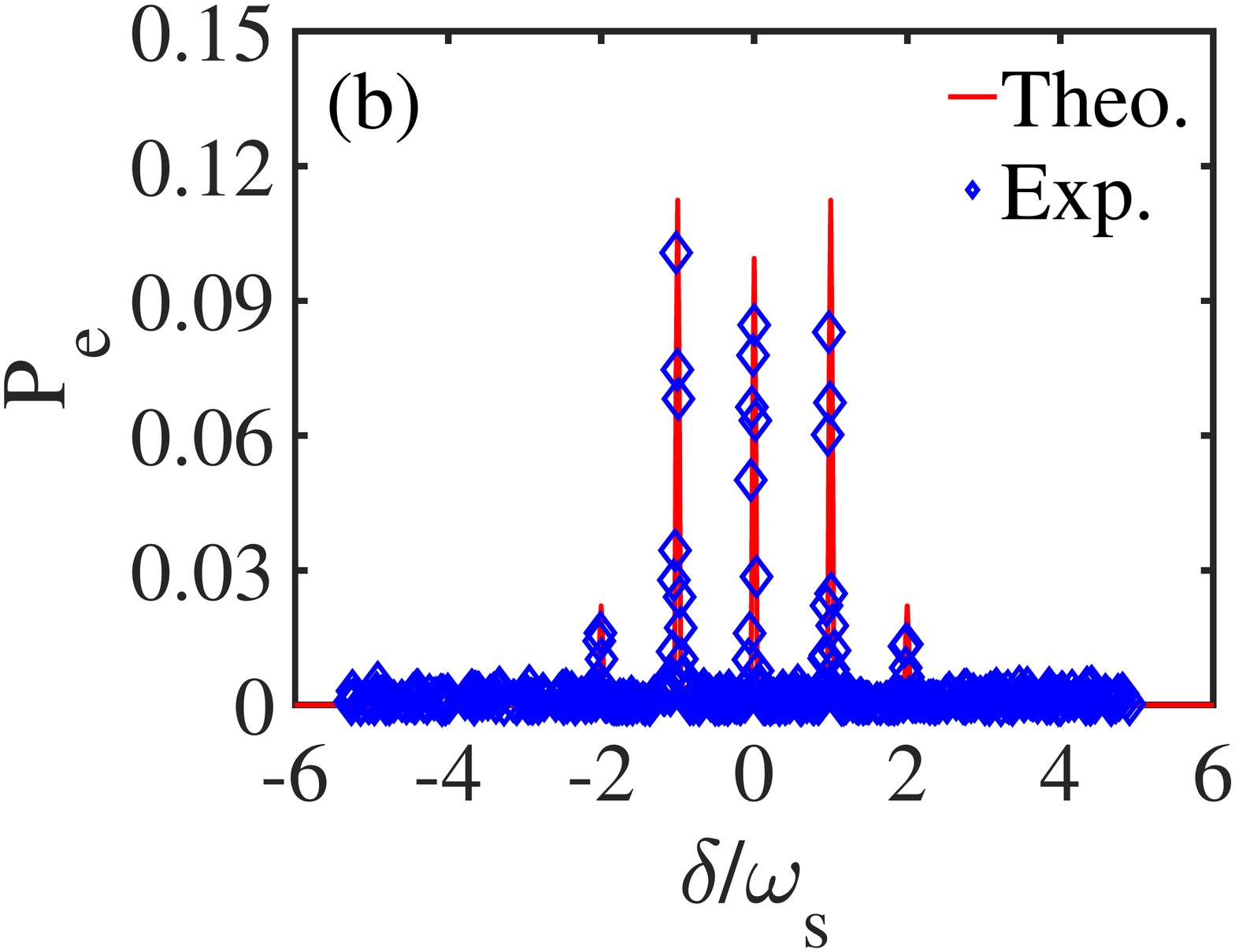}\\
	\includegraphics[width=0.5\linewidth]{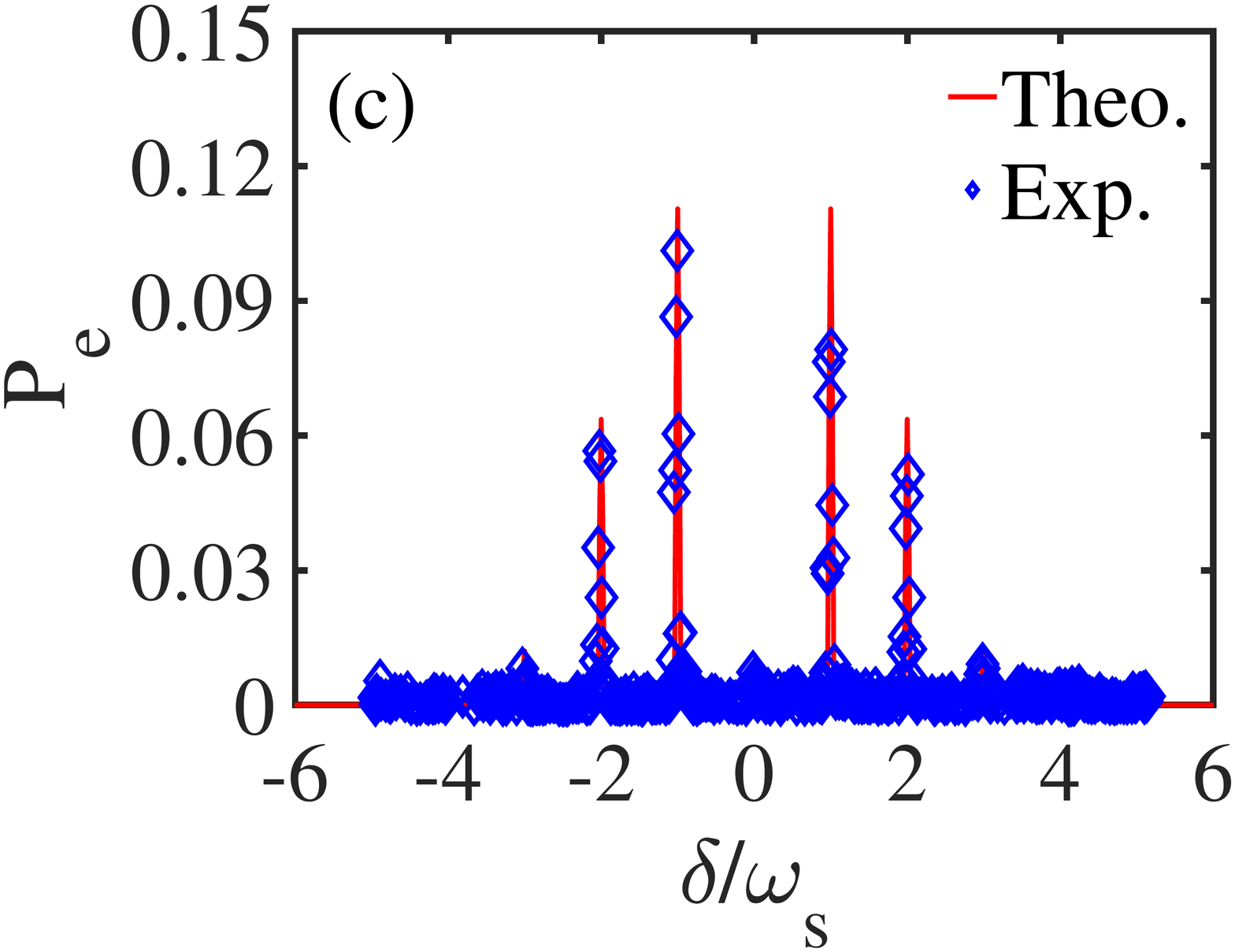}\hfill
	\includegraphics[width=0.5\linewidth]{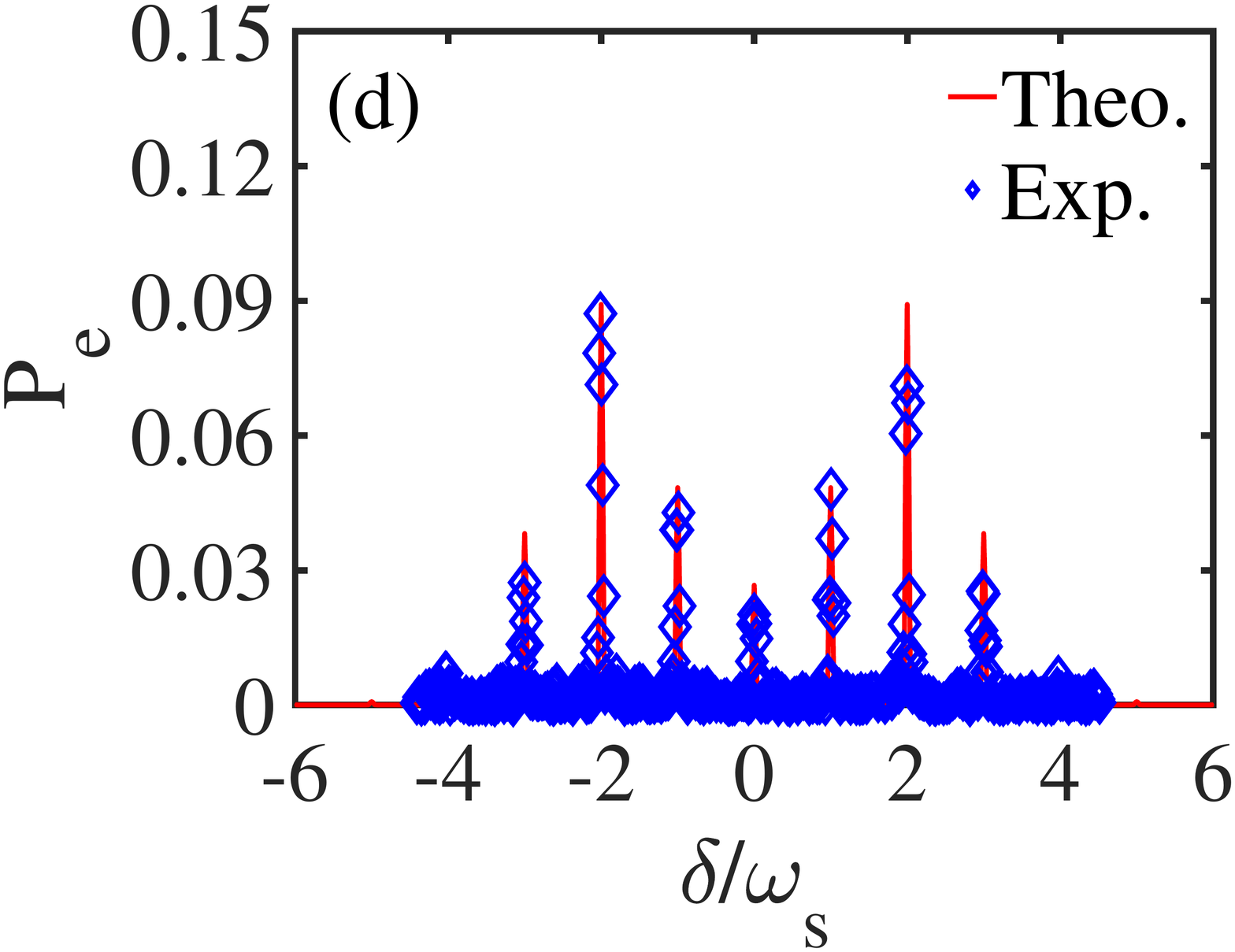}\\
	\includegraphics[width=0.5\linewidth]{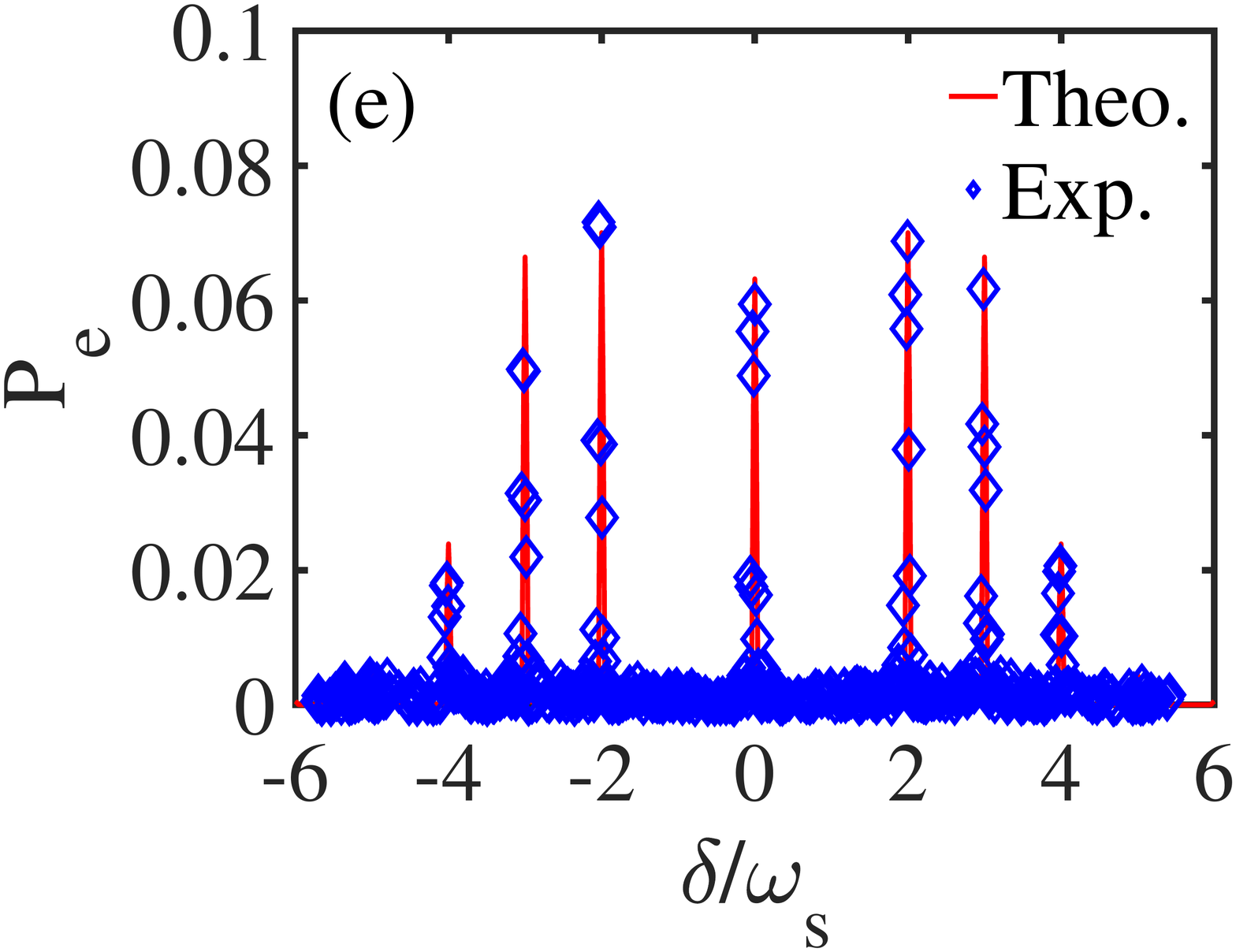}\hfill
	\includegraphics[width=0.5\linewidth]{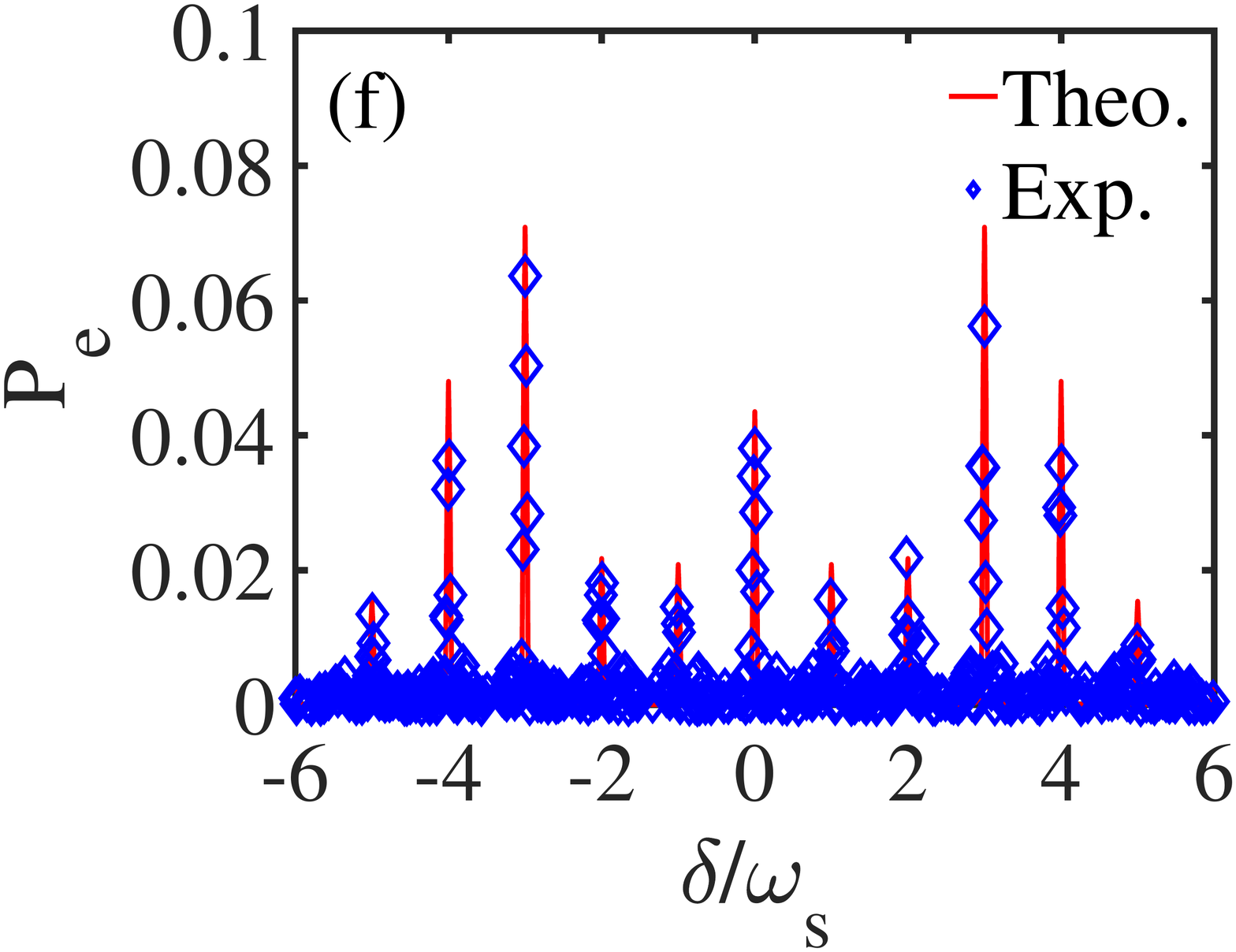}\\
	\caption{The Floquet degenerate spectrum at $150$ ms clock laser interrogation time, driving frequency $\omega_s/2\pi=200$ Hz, and residual magnetic field $B=6$ mG. The experimental data (blue diamonds) are compared to the theoretical results (red solid lines) at the driving voltages (a) $\bar{V}=0.5V$, (b) $\bar{V}=1.0V$, (c) $\bar{V}=1.5V$, (d) $\bar{V}=2.0V$, (e) $\bar{V}=2.5V$, (f) $\bar{V}=3.0V$, respectively.}%
	\label{Fig6}%
\end{figure}

\begin{figure}[t]
	\centering
	\includegraphics[width=1\linewidth]{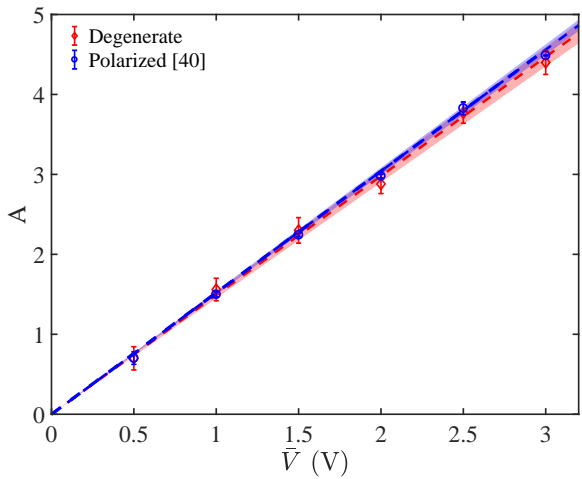}\\
	\caption{The relationship between renormalized driving amplitude $A$ and driving voltage $\bar{V}$ adding on the PZT in the Floquet degenerate system (red diamonds) and the Floquet polarized system (blue circles from Ref. \cite{YinCPL2021}). The dashed lines are linear fitting results, and the red color region show the 1$\sigma$ deviation from the fitting line of Floquet degenerate results.}%
	\label{Fig7}%
\end{figure}

\begin{figure}[t]
	\centering
	\includegraphics[width=0.5\linewidth]{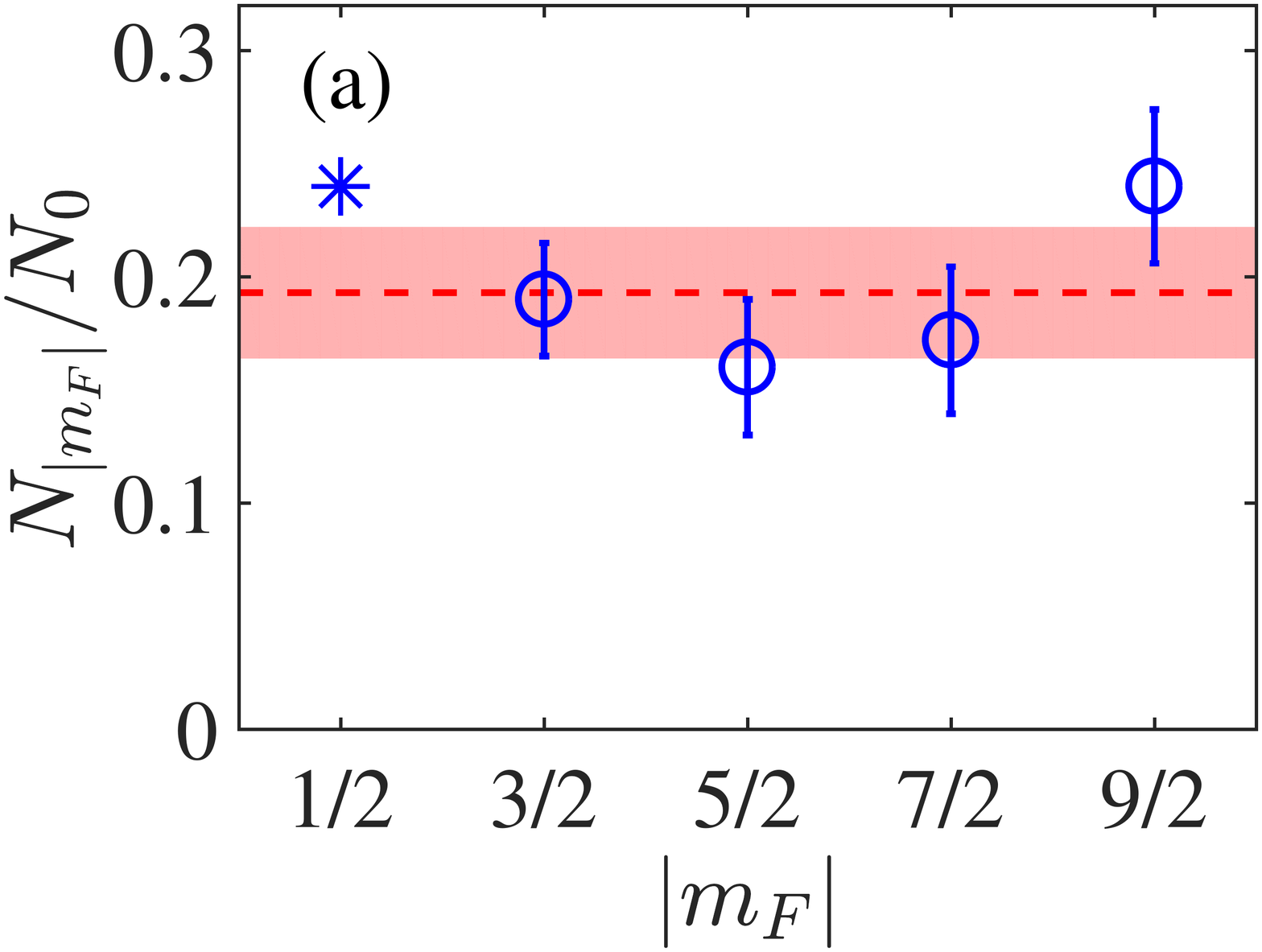}\hfill
	\includegraphics[width=0.5\linewidth]{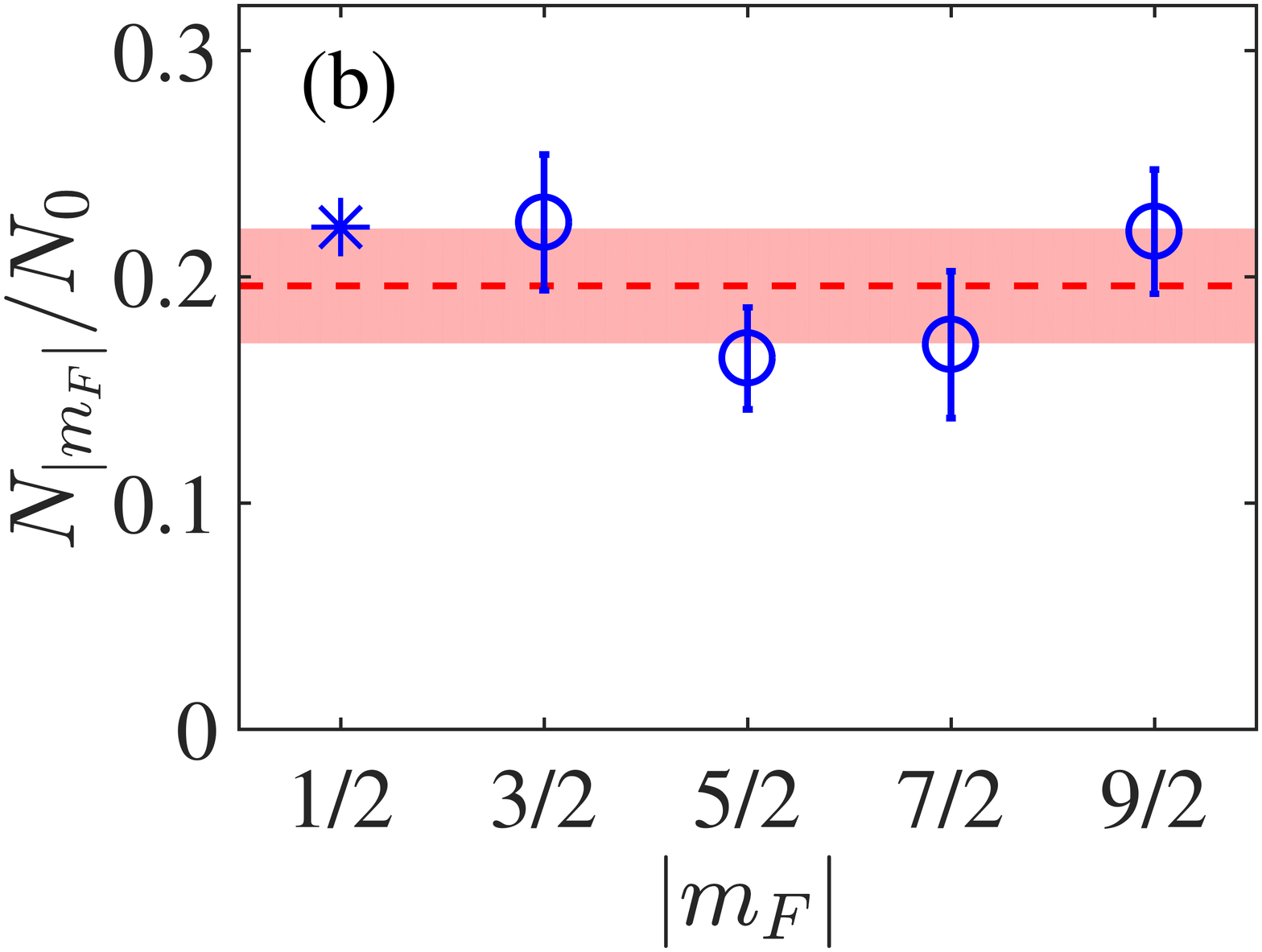}\\
	\includegraphics[width=0.5\linewidth]{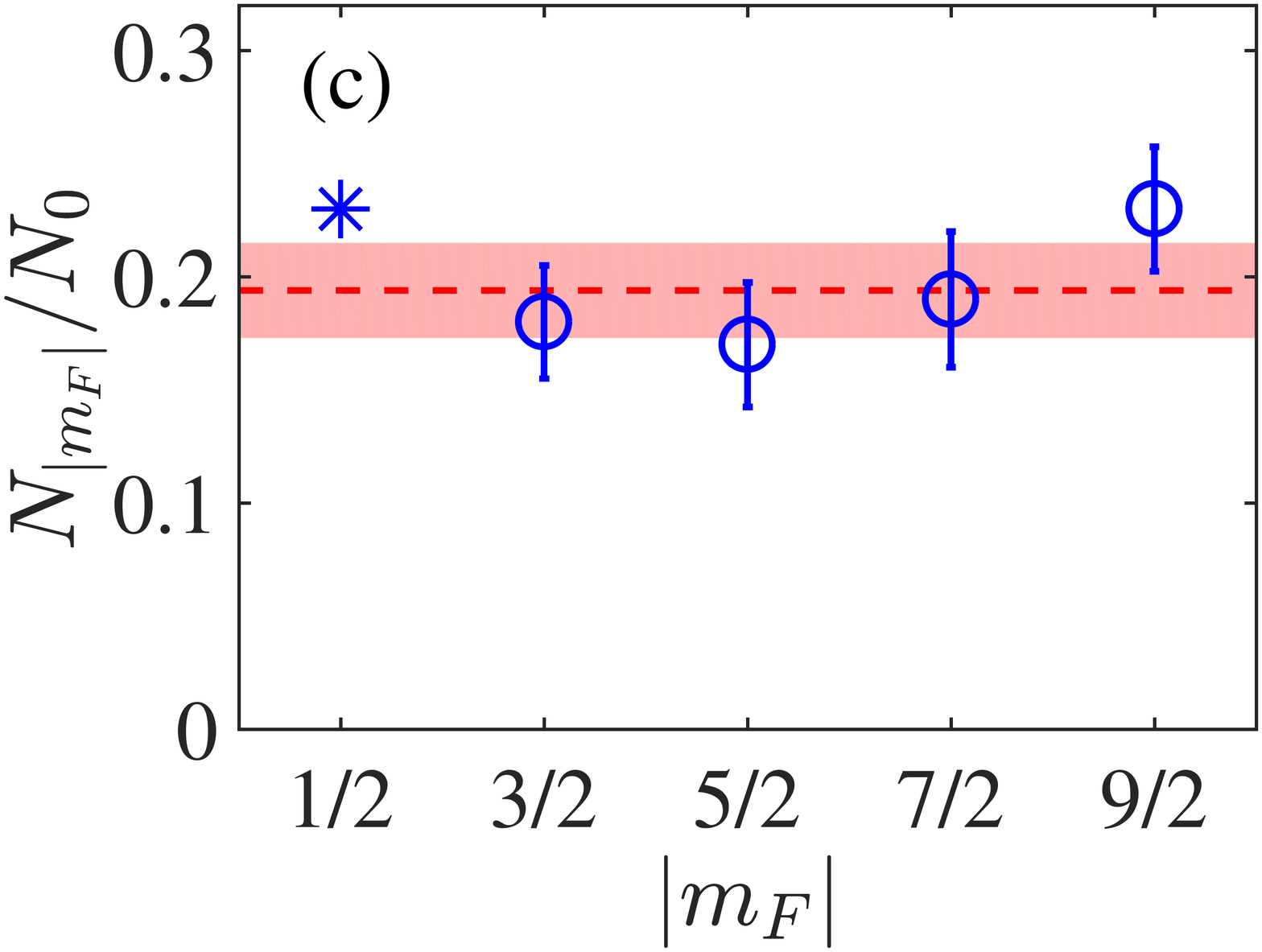}\hfill
	\includegraphics[width=0.5\linewidth]{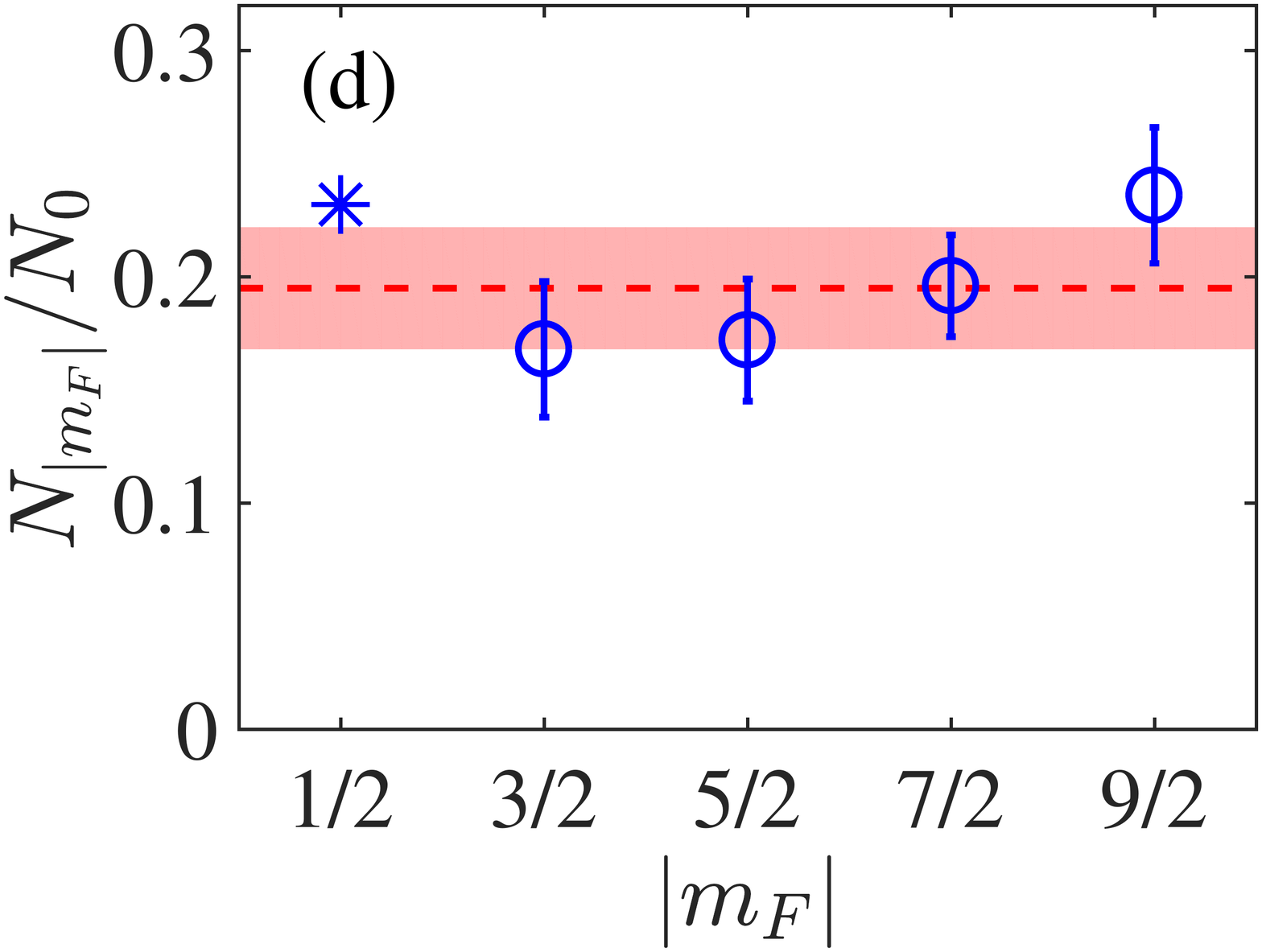}\\
	\includegraphics[width=0.5\linewidth]{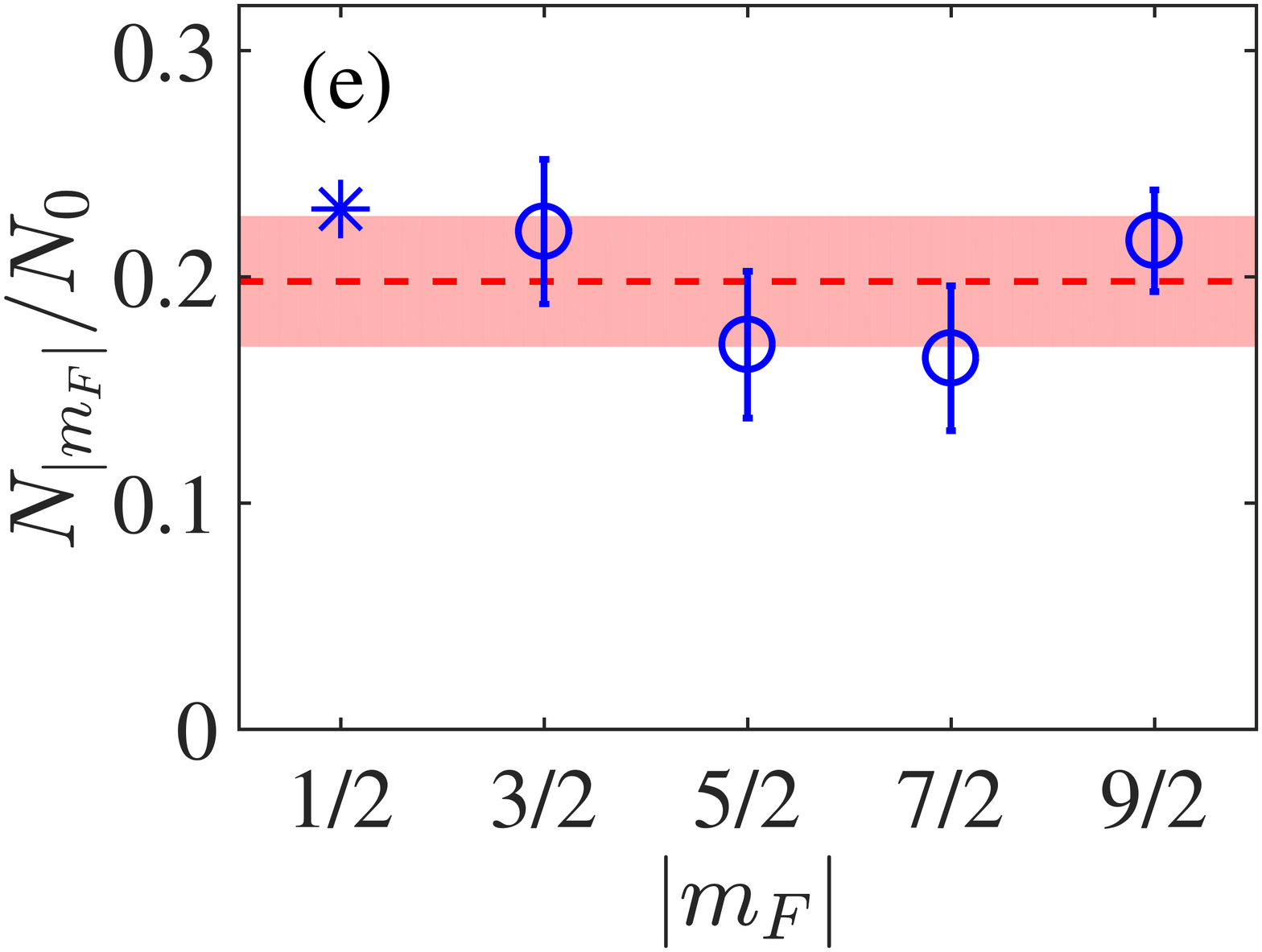}\hfill
	\includegraphics[width=0.5\linewidth]{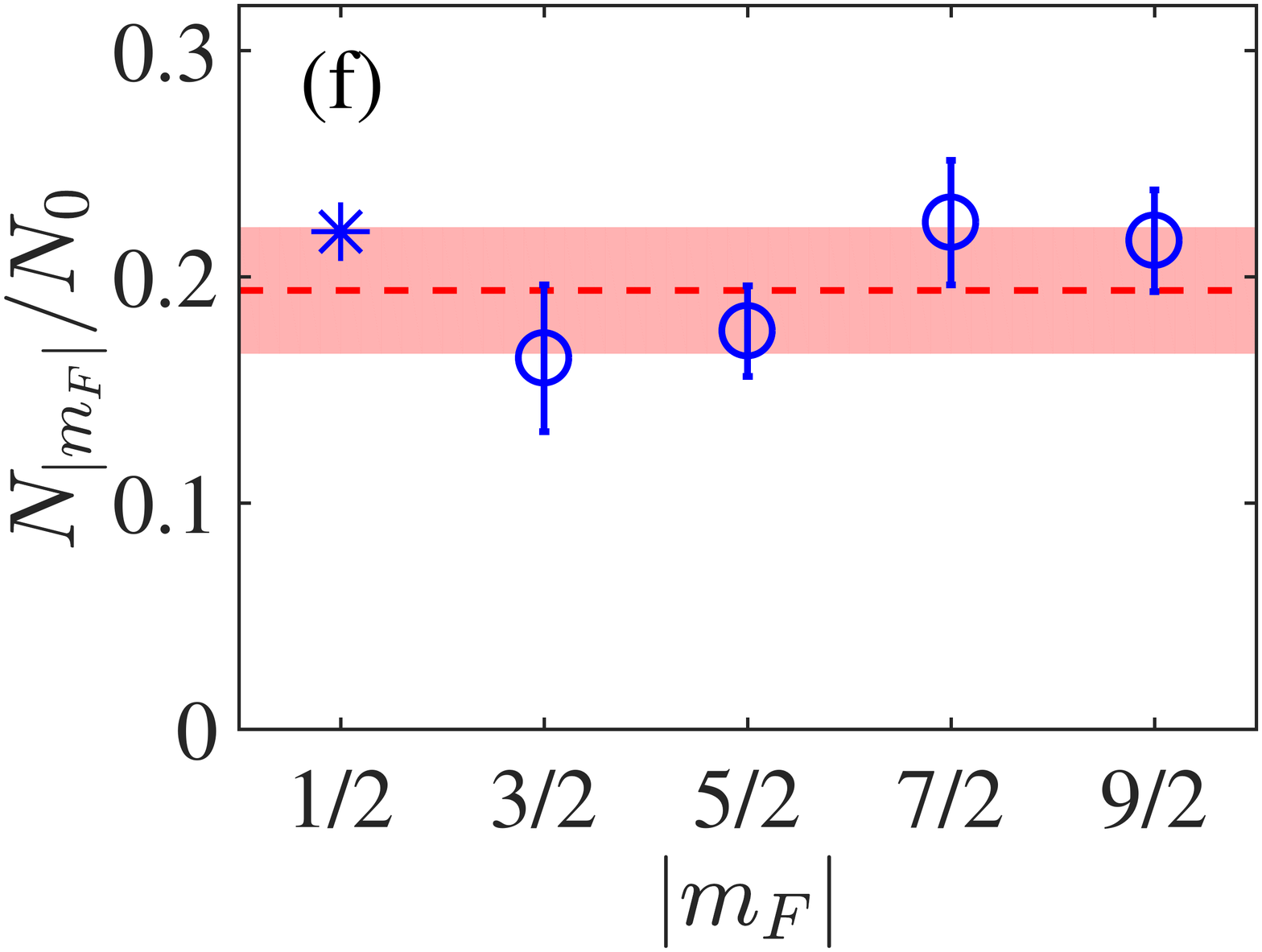}\\
	\caption{The atomic distribution at different sublevels obtained by fitting the experimental data with the Eq.(\ref{2}) from Floquet theory. The voltages added on the PZT are (a) $\bar{V}=0.5V$, (b) $\bar{V}=1.0V$, (c) $\bar{V}=1.5V$, (d) $\bar{V}=2.0V$, (e) $\bar{V}=2.5V$, (f) $\bar{V}=3.0V$, respectively. The dashed lines indicate the weighted mean, and the red region shapes the 1$\sigma$ standard deviation. The atomic distribution at sublevel $m_F=1/2$ labeled with star is calculated by subtracting the other sublevels from one.}%
	\label{Fig8}%
\end{figure}

Then, we turn to the atom distribution $N_{m_F}/N_0$. If the atoms are not uniform distributed, the height of all the Floquet sidebands will be strongly changed. In order to extract $N_{m_F}/N_0$, we experimentally measure the Rabi spectrum at different voltage and fit them according to Eq. (\ref{2}) by taking  $N_{m_F}/N_0$ and $A(\bar{V})$ as free parameters. Here, the contributions of $P_e$ at sublevel $m_F=\pm1/2$ are ignored because the corresponding Rabi frequency is very weak. In addition, $N_{m_F}/N_0$ is assumed not relevant to the sign of $m_F$. With this fitting method, we got the Rabi spectrum, see Fig. \ref{Fig6}; the relation between voltage and $A$, see Fig. \ref{Fig7}; and the atom distribution of all sublevels, see Fig. \ref{Fig8}. We will make concrete description about them in the following paragraphs. 

In the Fig. \ref{Fig6}, we show the experimental data of the degenerate Floquet Rabi spectroscopy compared with the theoretical results at different driving voltages. Generally, at all the driving voltages, the experimental results are in good agreement with the theoretical results. The intervals between the Floquet sidebands are same as the driving frequency $\omega_s$. Actually, there exists some small deviations because the Floquet spectra are obtained by scanning the clock laser with a step of $3$ Hz, so that it may fail to touch the peaks of all Floquet sidebands with the linewidths of a few Hz.

In our previous work \cite{YinCPL2021}, the relation between the driving amplitude $A$ and the voltage $\bar{V}$ adding on the PZT is quite linear. Before checking the uniformity of the atom distribution, it is better to verify the linearity between $A$ and $\bar{V}$, so that we can let $A$ linearly increase. The Fig. \ref{Fig7} strongly supports their linearity, and demonstrates the slope of degenerate case is very close to the polarized one (all atoms stay as sublevel $m_F=\pm 9/2$). Indeed, after linearly fitting the data $A=k\bar{V}$, the coefficient is $k=1.49(4)/V$ which is approximately same as polarized case $k=1.52/V$ \cite{YinCPL2021}.

Finally, as the most critical aspect, the atomic distributions extracted by fitting the experiment data are shown in Fig. \ref{Fig8}. Clearly, they are different and a little more atoms stay at some sublevels, such as $m_F=\pm9/2$. However, most of them fall into the 1$\sigma$ standard deviation of the weighted mean value (red region around $1/5$). The distribution in $m_F=\pm1/2$ is obtained by subtracting the atoms in other sublevels from unit. It is always large than $1/5$, which may result from the ignorance of the effect at $m_F=\pm1/2$ due to small Rabi frequency. Basing on these experimental results, we conclude that the periodic driving can not bring serious SU($N$) symmetry breaking.

\section{CONCLUSIONS AND DISCUSSIONS}

In this manuscript, we study the degenerate OLC system under the periodic modulation. Although the residual stray magnetic field can not be eliminated, we still can check whether the SU($N$) symmetry is broken by the FE. The driving amplitude and its relation with voltage adding on the PZT are not relevant to the Zeeman sublevels. Most importantly, the atomic population is nearly uniform within the standard error region.

Although the SU($N$) symmetry broken can not be totally ruled out, our experiment still support the periodic modulation of lattice frequency will not bring apparent influence on the SU($N$) symmetry. Our work will not only benefit the Floquet OLC \cite{ShallowYin2021}, but also shed a light on using the degenerate AEAs OLC for quantum computing and quantum simulation \cite{DaleyPRL2008,DaleyNIP2011}.

\section{ACKNOWLEDGMENTS}

This work is supported by the Special Foundation for theoretical physics Research Program of China (Grant No. 11647165), the China Postdoctoral Science Foundation Funded Project (Project No. 2020M673118) and the National Natural Science Foundation of China under Grant  No. 12147102. X.-F. Z. acknowledges funding from the National Science Foundation of China under Grants  No. 11874094, Fundamental Research Funds for the Central Universities Grant No. 2021CDJZYJH-003. W.-D.L. acknowledges the funding from the National Natural Science Foundation of China under Grant No. 11874247, the National Key Research and Development Program of China, Grant No. 2017YFA0304500, the Program of State Key Laboratory of Quantum Optics and Quantum Optics Devices, China, Grant No. KF201703, and the support from Guangdong Provincial Key Laboratory, Grant No. 2019B121203002.

\appendix

\section*{APPENDIX}
\label{apda}

\begin{figure}[htpb]
	\centering
	\includegraphics[width=0.5\linewidth]{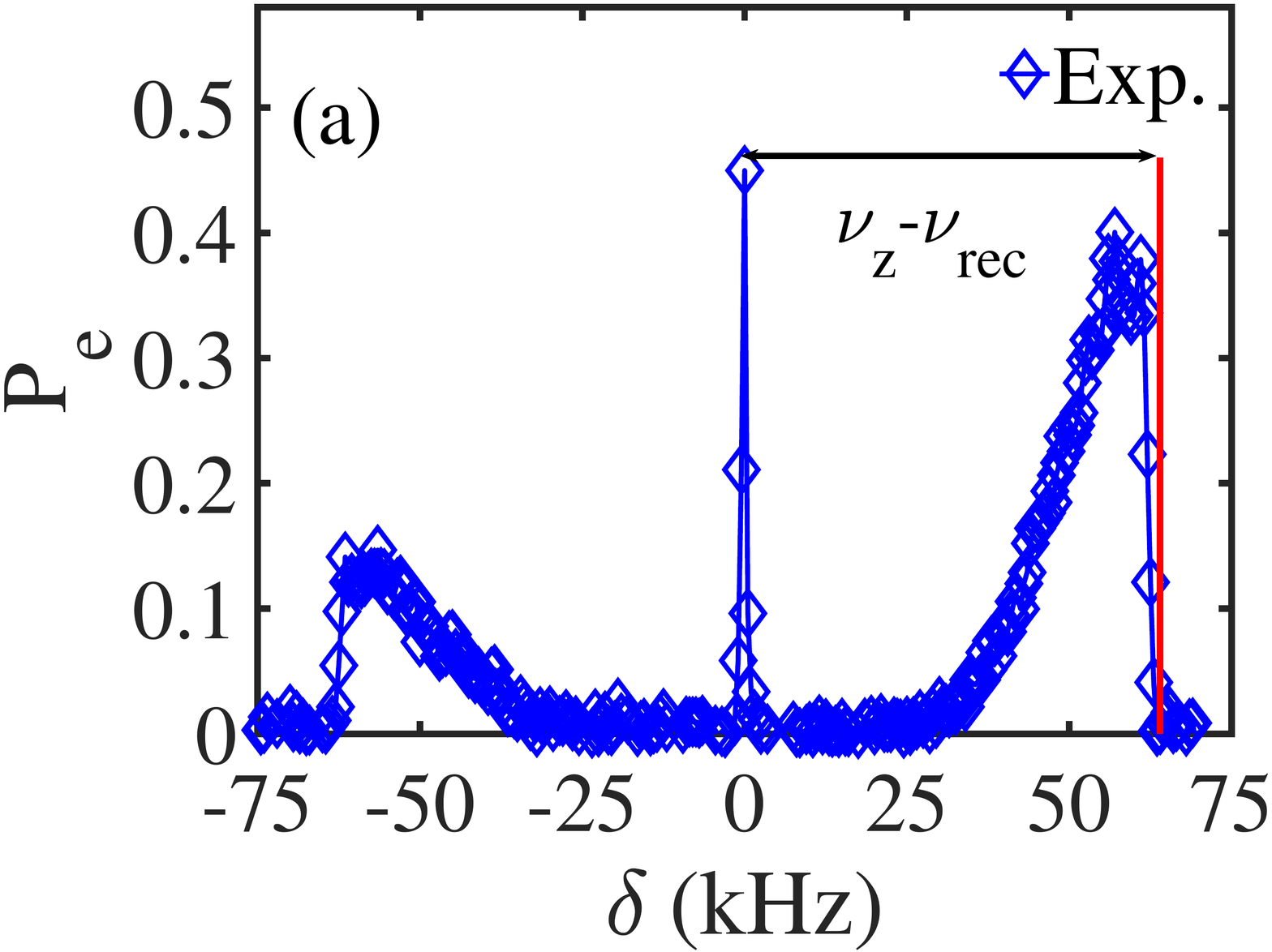}\hfill
	\includegraphics[width=0.5\linewidth]{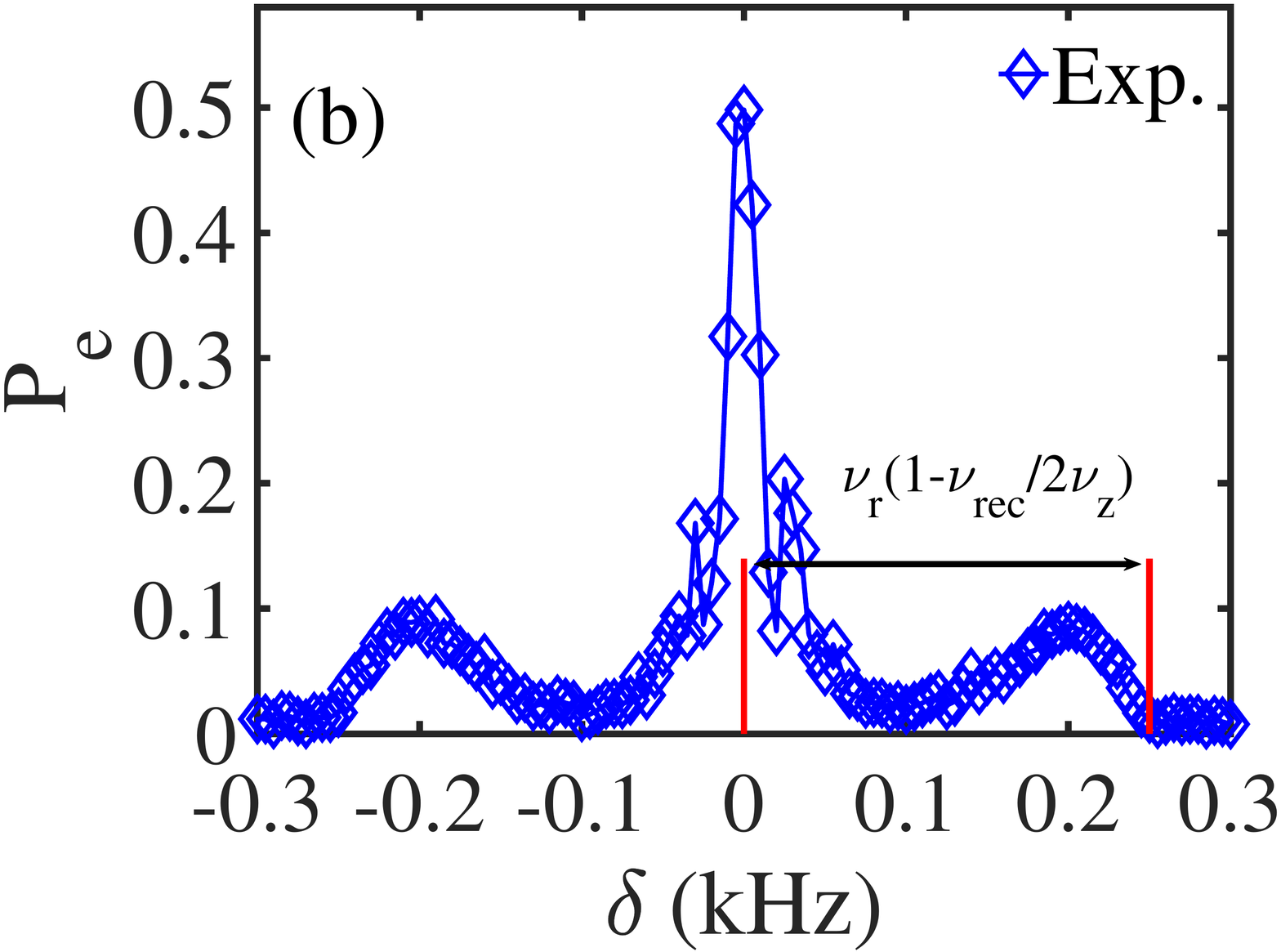}\\
	\includegraphics[width=0.5\linewidth]{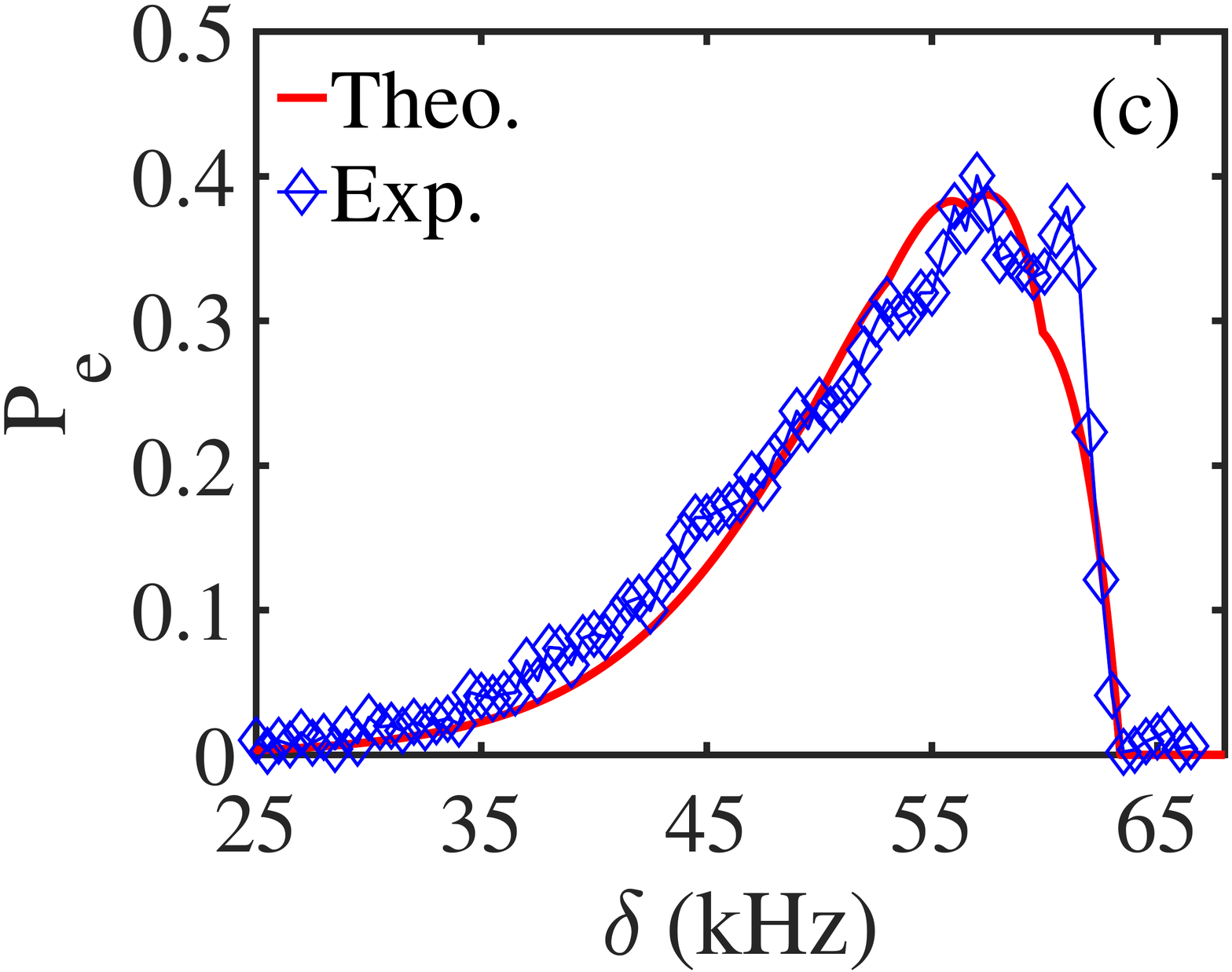}\\
	\caption{(Color online) Determination of the experimental parameters in non-driven case. (a). The experimental data of motional sideband spectrum in $z$ direction. (b). The experimental data of motional sideband spectrum in $r$ direction. (c). Theoretical fitting (red line) with experimental data (blue diamonds) for blue sideband in $z$ direction.}%
	\label{figs1}%
\end{figure}

Here we determine the experimental parameters (list in Table \ref{Table1}) using the motional sideband spectrum without driving (one can also refer to the Ref. \cite{BlattPRA2009} for an elaborate theoretical description). The motional trap frequencies [the longitudinal (transverse) trap frequency $\nu_z$ ($\nu_r$)], the number of motional states ($N_z$ and $N_r$), and the atom temperature ($T_z$ and $T_r$) are what we prepare to determine in this appendix. The motional sideband spectrum is obtained by changing the frequency of the clock laser with a step of $500$ Hz around the clock transition frequency, as shown in Fig. \ref{figs1}(a). The power of clock laser is about $1$ mW and the carrier peak is about $1$ kHz due to the saturation broadening. The complicated sideband spectrum indicates that we need a more subtle description of the energy spectrum more than a simple harmonic trap. By approximating the longitudinal potential as a $1$D harmonic trap with a quartic distortion and the transverse potential as a $2$D harmonic trap, we get the longitudinal blue-sideband energy gap which means the motional transition $|n_r,n_z\rangle \rightarrow |n_r,n_z+1\rangle$ \cite{BlattPRA2009}
\begin{equation}
\gamma(n_z)  = \nu_z-\nu_{\rm rec}(n_z+1)-\nu_{\rm rec}\frac{\nu_r}{\nu_z}(n_r+1).  \label{s8}
\end{equation}

Similarly, we get the transverse motional sideband spectrum when a slight angle between the clock and lattice laser beam. And the power of clock laser is reduced to about $10 \mu$W to get a transverse resolved sideband spectrum shown in Fig. \ref{figs1}(b). So the transverse blue sideband energy gap which means the motional transition $|n_r,n_z\rangle \rightarrow |n_r+1,n_z\rangle$ is
\begin{equation}
\gamma(n_r)  = \nu_r-\nu_{\rm rec}\frac{\nu_r}{\nu_z}(n_z+1/2).  \label{s9}
\end{equation}

The position of longitudinal blue sideband sharp edge means the largest energy gap which determines the longitudinal trapping frequency $\nu_z$ by $\gamma(n_z)(n_z=0,n_r=0) \simeq \nu_z-\nu_{\rm rec}$. similarly, we can determine the transverse trapping frequency by $\gamma(n_r)(n_z=0)=\nu_r-\nu_{\rm rec}\nu_r/2\nu_z$. Thus we can read from Fig. \ref{figs1}(a) and Fig. \ref{figs1}(b) that $\nu_z=66.8$ kHz and $\nu_r=250$ Hz. Then we can determine the trap depth and the beam waist by $U_0=\nu_z^2E_R/4\nu_{\rm rec}^2 \simeq 94$ $E_R$, $w_0=\sqrt{U_0/m\nu_r^2\pi^2}=49$ $\mu$m, and number of motional states $N_z\simeq U_0/h\nu_z=5$, $N_r\simeq U_0/h\nu_r=1336$. Now we determine the atom temperature. The longitudinal red-sideband means the motional transition $|n_r,n_z\rangle \rightarrow |n_r,n_z-1\rangle$. If we regardless of the details of the sideband line shapes, the only difference between the blue and red sidebands is that the Boltzmann weights are shifted according to $n_z \rightarrow n_z+1$. And the longitudinal ground state $n_z=0$ dose not contribute to the red sideband, so we can determine the longitudinal temperature with the ratio of sidebands cross sections
\begin{equation}
\frac{\sigma_{\rm red}}{\sigma_{\rm blue}}=\frac{\sum_{n_z=1}^{N_z} q(n_z)}{\sum_{n_z=0}^{N_z-1}q(n_z)},  \label{s10}
\end{equation}
which give us the longitudinal temperature $T_z=3.5$ $\mu$K. The transverse temperature can be extracted from the longitudinal blue-sideband line shape with equation \cite{BlattPRA2009}
\begin{equation}
\sigma_{\rm blue}(\delta) \propto \sum_{n_z=0}^{N_z-1}q(n_z)\frac{\alpha\tilde{\delta}}{\tilde{\gamma}(n_z)}e^{-\alpha\tilde{\delta}}\Theta[\tilde{\gamma}(n_z)\tilde{\delta}],
\label{s11}
\end{equation}
where $\alpha=[\tilde{\gamma}(n_z)/\nu_{rec}](h\nu_z/k_BT_r)$, $\tilde{\gamma}(n_z)=\nu_z-\nu_{\rm rec}(n_z+1)$, $\tilde{\delta}=1-\delta/\tilde{\gamma}(n_z)$, $\Theta$ is the Heaviside function. By fitting with experimental data (Fig. \ref{figs1}(c)), we get the transverse temperature $T_r=4.0$ $\mu$K.

\bibliographystyle{apsrev4-1}
\bibliography{degenerate_15}

\end{document}